\documentclass[a4paper,12pt]{article}
\usepackage[a4paper,text={16.8cm,22.4cm}]{geometry}
\usepackage{amsmath,amssymb,latexsym,bm,braket}
\usepackage{cite}
\usepackage{graphicx}
\usepackage{color}

\allowdisplaybreaks 
\newcommand{\Tr}{\mathrm{Tr}}
\newcommand{\Li}{\mathrm{Li}}
\newcommand{\ff}{f\hspace{-0.4em}f}
\newcommand{\fff}{f\hspace{-0.2em}f}
\newcommand{\DD}{D\hspace{-0.7em}D}
\newcommand{\dd}{d\hspace{-0.4em}d}

\newcommand{\be}{\begin{equation}}
\newcommand{\ee}{\end{equation}}
\newcommand{\bea}{\begin{eqnarray}}
\newcommand{\eea}{\end{eqnarray}}
\newcommand{\cusp}{\mbox{{\tiny cusp}}}

\begin{document}

\begin{titlepage}

  \begin{flushright}
    MZ-TH/12-19 \\
    ZU-TH 05/12 \\
    May 17, 2012
  \end{flushright}
  
  \vspace{5ex}
  
  \begin{center}
    \textbf{ \Large Soft-gluon resummation for boosted top-quark
      production at hadron colliders} \vspace{7ex}
    
    \textsc{Andrea Ferroglia$^a$, Ben D. Pecjak$^b$, and Li Lin
      Yang$^c$} \vspace{2ex}
  
    \textsl{${}^a$New York City College of Technology, 300 Jay Street\\
      Brooklyn, NY 11201, USA
      \\[0.3cm]
      ${}^b$Institut f\"ur Physik (THEP), Johannes Gutenberg-Universit\"at\\
      D-55099 Mainz, Germany
      \\[0.3cm]
      ${}^c$Institute for Theoretical Physics, University of Z\"urich\\
      CH-8057 Z\"urich, Switzerland}
  \end{center}

  \vspace{4ex}

  \begin{abstract}
   
    We investigate the production of highly energetic top-quark pairs
    at hadron colliders, focusing on the case where the invariant mass
    of the pair is much larger than the mass of the top quark. In
    particular, we set up a factorization formalism appropriate for
    describing the differential partonic cross section in the double
    soft and small-mass limit, and explain how to resum simultaneously
    logarithmic corrections arising from soft gluon emission and from
    the ratio of the pair-invariant mass to that of the top quark to
    next-to-next-to-leading logarithmic accuracy. We explore the
    implications of our results on approximate next-to-next-to-leading
    order formulas for the differential cross section in the soft
    limit, pointing out that they offer a simplified calculational
    procedure for determining the currently unknown delta-function
    terms in the limit of high invariant mass.

  \end{abstract}

\end{titlepage}

\section{Introduction}
\label{sec:intro}

Top-quark pair production plays an important role in the physics
programs of hadron colliders such as the Tevatron and the LHC. While
much information about the top quark is already available, its high
production rate at the LHC will eventually bring studies of its
properties into the realm of precision physics. An especially useful
observable is the differential cross section at large values of the
top-pair invariant mass $M$. Many models of physics beyond the
Standard Model predict the existence of new particles which decay into
energetic top quarks and whose characteristic signal would be either
resonant bumps or more subtle distortions in the high invariant mass
region of the differential distribution. In fact, Tevatron
measurements of the forward-backward asymmetry at high values of the
pair invariant mass may already hint at the existence of such
particles \cite{Aaltonen:2011kc, Abazov:2011rq}. Therefore, precision
calculations of the pair invariant-mass distribution within the
Standard Model are well motivated.

The starting point for any study of two-particle inclusive
differential cross sections such as the pair invariant-mass
distribution is the next-to-leading order (NLO) calculations carried
out roughly two decades ago \cite{Mangano:1991jk}. Recently, these
have been supplemented with soft-gluon resummation at
next-to-next-to-leading logarithmic (NNLL) accuracy in
\cite{Ahrens:2009uz, Ahrens:2010zv}, building on the next-to-leading
logarithmic (NLL) results of \cite{Almeida:2008ug}. While such
resummed calculations contain what are argued to be the dominant
perturbative corrections at next-to-next-to-leading order (NNLO) and
beyond, they suffer from two potential shortcomings. First, while they
fully determine the coefficients of a subset of logarithmic
plus-distribution corrections which become large in the soft limit
$z=M^2/\hat{s}\to 1$, with $\sqrt{\hat{s}}$ the partonic
center-of-mass energy, they do not fully determine the delta-function
corrections in this limit at NNLO. The numerical contribution of these
unknown coefficients as well as less singular terms in the soft limit
are typically estimated through the method of scale variations, but
this is by no means a fail-safe technique and additional information
about the structure of these terms is valuable. Second, while
\cite{Ahrens:2010zv} uses the parametric counting $M \sim m_t$, when
the top quarks are truly very boosted this counting breaks down. One
must assume instead that $M \gg m_t$, and recognize that resummed
perturbation theory should also take into account powers of mass
logarithms of the ratio $m_t/M$.

The primary aim of this work is to develop the theoretical framework
needed to describe the invariant-mass distribution in the double soft
and small-mass limit, and to explore some of its implications for
perturbative predictions for the large-$M$ distribution. The basic
idea behind our approach is to weave together current understanding of
factorization in either the small-mass or the soft limit into a
unified description encompassing both. The first component is the
factorization of partonic cross sections for highly boosted
heavy-quark production worked out in \cite{Mele:1990cw}. In the case
at hand, where $m_t \ll M$, the results of that work imply that the
partonic cross section can be factorized into a convolution of two
functions: the cross section for {\it massless} quark production, and
a convolution perturbative fragmentation functions for each of the
heavy quarks. Given this form of the cross section for the small-mass
limit, it is then an easy matter to perform the additional layer of
factorization for the soft limit on the component parts. On the one
hand, the massless partonic cross section in this limit can be
factorized into a product of soft and hard functions using the
techniques from \cite{Kidonakis:1997gm}, and on the other hand, the
fragmentation function can be factorized into a product of collinear
and soft-collinear functions using the results of
\cite{Korchemsky:1992xv, Cacciari:2001cw, Gardi:2005yi,
  Neubert:2007je}. A fully resummed cross section appropriate for both
limits is then obtained by deriving and solving the
renormalization-group (RG) evolution equations for the different
functions separately. The anomalous dimensions appearing in the RG
equations are known to the level sufficient for resummation of both
mass and soft logarithms to NNLL accuracy. As simple and obvious as
this approach is, it has yet to be fully worked out for any particular
observable in top-quark pair production at hadron collider
experiments.

This formalism for the simultaneous resummation of soft and mass
logarithms in the invariant-mass distribution is interesting in its
own right. Moreover, with use of a proper matching procedure, it
provides supplemental information to the current state-of-the-art
predictions based on soft-gluon resummation with the counting $m_t\sim
M$ \cite{Ahrens:2010zv}. Particularly important in this regard is its
use as a tool to calculate, up to easily quantifiable power
corrections in $m_t/M$, the full NNLO corrections to the massive hard
and soft functions. Together, these pieces determine the coefficient
of the delta-function coefficient in the fixed-order expansion at
NNLO, a missing piece in currently available ``approximate NNLO''
formulas for generic values of the top-quark mass. Using our
factorization formula for the double soft and small-mass limit, we can
calculate the pieces of the NNLO delta-function correction enhanced by
logarithms of the ratio $m_t/M$. Furthermore, using the explicit NNLO
results for the heavy-quark fragmentation function
\cite{Melnikov:2004bm} and the virtual corrections to massless
$q\bar{q} \to q'\bar{q}'$ \cite{Anastasiou:2000kg} and $gg \to
q\bar{q}$ \cite{Anastasiou:2001sv} scattering, we can very nearly
determine the piece of the delta-function coefficients which is
constant in the limit $m_t/M \to 0$. The missing piece is the NNLO
soft function for massless partons, related to double real emission
for $gg \to q\bar{q}$ and $q\bar{q} \to q\bar{q}$ scattering in the
soft limit. We do not calculate this function here, but plan to return
to it in future work. While these delta-function pieces of the NNLO
partonic cross section are of N$^3$LL in the counting of soft-gluon
resummation, including them can only make the predictions more precise
and potentially strengthen the arguments in favor of the logarithmic
counting underlying approximate NNLO formulas. This is currently an
open point in soft-gluon resummation for the invariant-mass
distribution, where the assumed dominance of logarithmic corrections
is justified mainly through numerical studies at NLO and arguments
based on dynamical threshold enhancement \cite{Becher:2007ty}.

The remainder of the paper is organized as follows. First, in
Section~\ref{sec:factorization}, we derive a factorization formula for
the partonic cross section valid in the double small-mass and soft
limit. This fixed-order expression contains large logarithms for any
choice of the factorization scale. We deal with this in
Section~\ref{sec:resummation} by deriving and solving the RG equations
for the component functions of the factorization formula, presenting
in addition explicit perturbative results for their fixed-order
expansions. In Section~\ref{sec:rescs} we combine those results into
an expression for the resummed partonic cross section at NNLL in
perturbation theory, and discuss different matching procedures needed
to take into account power-suppressed terms away from the double
small-mass and soft limit. In that section we also discuss approximate
NNLO implementations of the NNLL formula. Finally, in
Section~\ref{sec:numerics}, we make preliminary explorations into
phenomenological consequences of our results. We conclude in
Section~\ref{sec:conclusions}.

\section{Factorization in the soft and small-mass limits}
\label{sec:factorization}

We study the top-quark pair production process
\begin{align}
  N_1(P_1) + N_2(P_2) \rightarrow t(p_3) +\bar{t}(p_4) + X(p_X) \, ,
\end{align}
where $N_1$ and $N_2$ are the colliding protons (or proton-antiproton
pair), $X$ is an inclusive hadronic state, and the top quarks are
treated as on-shell particles. Two partonic channels contribute at
lowest order in perturbation theory: the quark annihilation channel
\begin{align}
  q(p_1) + \bar{q}(p_2) \rightarrow t(p_3) +\bar{t}(p_4) \, ,
\end{align}
and the gluon fusion channel
\begin{align}
  g(p_1) + g(p_2) \rightarrow t(p_3) +\bar{t}(p_4) \, .
\end{align}
The momenta of the incoming partons are related to the hadron momenta
by the relation $p_i = x_i P_i$ ($i=1,2$). The relevant Mandelstam
invariants are defined as
\begin{gather}
  s = (P_1+P_2)^2 \, , \quad \hat{s} = (p_1+p_2)^2 \, , \quad M^2 =
  (p_3+p_4)^2 \, , \nonumber
  \\
  t_1 = (p_1-p_3)^2 - m_t^2 \, , \quad u_1 = (p_2-p_3)^2 -m_t^2 \, .
\end{gather}
In order to describe the invariant-mass distribution near the partonic
threshold, it is convenient to introduce the following variables:
\begin{align}
  z = \frac{M^2}{\hat{s}} \, , \quad \tau = \frac{M^2}{s} \, , \quad
  \beta_t = \sqrt{1-\frac{4m_t^2}{M^2}} \, , \quad \beta=
  \sqrt{1-\frac{4m_t^2}{\hat{s}}} \, .
\end{align}
The quantity $\beta_t$ is the 3-velocity of the top quarks in the
$t\bar{t}$ rest frame. In the soft limit $z \to 1$, one has $\beta \to
\beta_t$. Moreover, in that limit the scattering angle $\theta$ is
related to the Mandelstam variables according to
\begin{align}
  \label{eq:tu}
  t_1 = -\frac{M^2}{2} ( 1 - \beta_t \cos\theta ) \, , \qquad u_1 =
  -\frac{M^2}{2} ( 1 + \beta_t \cos\theta ) \, ,
\end{align}
and $M^2+t_1+u_1=0$ can be used to eliminate $u_1$ as an independent
variable.

We will be interested in the double differential cross section with
respect to the invariant mass of the top-quark pair and the scattering
angle $\theta$ in the parton center-of-mass frame. According to
factorization in QCD, the double differential cross section can be
written as a convolution of a partonic cross section with parton
distribution functions (PDFs). We write this as
\begin{align}
  \label{eq:genfact}
  \frac{d^2\sigma}{dMd\cos\theta} = \frac{8\pi\beta_t}{3sM} \sum_{i,j}
  \int_\tau^1 \frac{dz}{z} \, \ff_{ij}(\tau/z,\mu_f) \,
  C_{ij}(z,M,m_t,\cos\theta,\mu_f) \, ,
\end{align}
where $\mu_f$ is the factorization scale. The parton luminosity
functions functions $\ff_{ij}$ are defined as a convolution of PDFs:
\begin{align}
\label{eq:lumdef}
  \ff_{ij}(y,\mu_f) = f_{i/N_1}(y,\mu_f) \otimes f_{j/N_2}(y,\mu_f) \,
  .
\end{align}
Here and throughout the paper the symbol $\otimes$ denotes the
following convolution between two functions
\begin{align}
  f(z) \otimes g(z) = \int_{z}^1 \frac{dx}{x} \, f(x) \, g(z/x) \, .
\end{align}
When there are several arguments in the functions $f$ and $g$, the
convolution is always over the first argument. As described in more
detail below, we choose to define the PDFs with $n_l=5$ active light
flavors, so that all physics associated with the scale of the
top-quark is absorbed into the perturbative coefficient functions
$C_{ij}$. These coefficient functions are proportional to differential
partonic cross sections. Our aim is to study the factorization
properties of these partonic cross sections in the double soft and
small-mass limit, where $(1-z) \ll 1$ and $m_t \ll M$.\footnote{More
  precisely, we work in the limit where the Mandelstam variables
  $\hat{s},t_1,u_1\gg m_t$.} Our strategy is to first discuss the soft
and small-mass limits separately, and then combine them into a single
formula which is true for both limits simultaneously.

Factorization of differential partonic cross sections in the soft
limit has been studied in \cite{Kidonakis:1996aq, Kidonakis:1997gm,
  Banfi:2005cv, Kidonakis:2000ui, Kidonakis:2001nj, Kidonakis:2003qe,
  Kidonakis:2008mu, Kidonakis:2011zn, Kidonakis:2010dk, Ahrens:2010zv,
  Ahrens:2011mw}. In the soft limit $z \to 1$, the partonic cross
section can be factorized into a hard function and a soft function
according to
\begin{align}
  \label{eq:softfact}
  C_{ij}(z,M,m_t,\cos\theta,\mu_f) &= \Tr \left[
    \bm{H}_{ij}^m(M,m_t,t_1,\mu_f) \, \bm{S}_{ij}^m
    (\sqrt{\hat{s}}(1-z),m_t,t_1,\mu_f) \right] + \mathcal{O}(1-z) \,
  ,
\end{align}
where we have used that in the soft limit dependence on the scattering
angle can be expressed in terms of $t_1$, see (\ref{eq:tu}). The
superscript $m$ on the hard function $\bm{H}_{ij}^m$ and the soft
function $\bm{S}_{ij}^m$ indicates that they are computed with finite
top-quark mass, as opposed to the massless hard and soft functions
introduced below. Both of these functions are matrices in the space of
color-singlet operators for Born-level production. The hard function
is related to virtual corrections to the two-to-two scattering
processes $q\bar{q} (gg) \to t\bar{t}$. The soft function is related
to real emissions in the soft limit, or more precisely to the vacuum
expectation value of a Wilson-loop operator built from time and
light-like Wilson lines associated with soft emissions from the heavy
and light quarks. Note that in this limit, we only need to consider
$ij=q\bar{q},gg$, since the $qg$ channel is suppressed by powers of
$(1-z)$.

Factorization of partonic cross sections for heavy-quark production in
the small-mass limit was considered in \cite{Mele:1990cw}. It was
shown that partonic cross sections in this limit can be factorized
into a product of massless cross sections with perturbatively
calculable heavy-quark fragmentation functions. Generically, for a
cross section differential in the energy fraction $z= E/E_{\rm max}$
of the top quark, this factorization is written as (see, for instance,
\cite{Melnikov:2004bm})
\begin{align}
  \label{eq:fragdef}
  \frac{d \sigma_t}{dz}(z,m_t,\mu) = \sum_a \int_z^1 \frac{dx}{x}
  \frac{d \tilde{\sigma}_a}{dx}(x,m_t,\mu) D_{a/t}^{(n_l+n_h)}
  \left(\frac{z}{x},m_t,\mu\right) \, ,
\end{align}
where $d \tilde{\sigma}_a/ dx$ is the
$\overline{\text{MS}}$-renormalized differential cross section for the
production of a massless parton $a$, and $D_{a/t}^{(n_f)}$ is the
heavy-quark fragmentation function defined using an $\alpha_s$ with
$n_f$ active flavors. The sum over the massless partons labeled by $a$
includes the case $a=t$, and the heavy quark is considered massless in
the calculation of $d\tilde{\sigma}_t$. Much as the PDFs describe
radiation collinear to initial-state partons, the heavy-quark
fragmentation functions describe radiation collinear to the energetic
final-state heavy quarks. The heavy-quark fragmentation functions are
however perturbatively calculable, since the mass of the top quark
serves as a collinear regulator.

In the case of the invariant-mass distribution in top-quark pair
production at hadron colliders, we need to modify the generic formula
(\ref{eq:fragdef}) in several ways. First, since this observable
contains information about both the top and the anti-top quark, we
need a fragmentation function for each of them. Second, we must
introduce heavy-flavor coefficients related to matching six-flavor
PDFs onto five-flavor ones, which induces an additional source of
$m_t$-dependence into the formula. Finally, although not strictly
necessary, we will follow \cite{Neubert:2007je} and also perform such
a matching for the fragmentation functions. The matching relations
between the PDFs and fragmentation functions in the $n_l+n_h$- and
$n_l$-flavor theories are
\begin{align}
  \label{eq:decoupling}
  D_{a/t}^{(n_l+n_h)}(z,m_t,\mu_f) & = C_{a/t}(z,m_t,\mu_f) \otimes
  D^{(n_l)}_{t/t}(z,m_t,\mu_f) \, ,
  \\
  \ff_{ij}^{(n_l+n_h)}(z,m_t,\mu_f)& = C^{ij}_{\fff}(z,m_t,\mu_f)
  \otimes \ff_{ij}^{(n_l)}(z,\mu_f) \, .
\end{align}
The heavy-flavor matching coefficients $C_{a/t}$ and $C^{ij}_{\fff}$
on the right-hand side of the above equation are proportional to
powers of $n_h=1$. They are obtained by comparing partonic matrix
elements with and without the top quark as an active flavor, and are
known to NNLO in fixed-order for both the fragmentation functions
\cite{Neubert:2007je} and the parton luminosity functions
\cite{Chuvakin:2001ge}. We will encounter them again in
Section~\ref{sec:rescs}, when we discuss the RG running of massless
coefficient functions to scales below the flavor threshold at
$\mu_t\sim m_t$.

Taking these points into account, the factorization formula for the
coefficient function (\ref{eq:genfact}) in the small-mass limit
reads\footnote{We have used a slight abuse of notation and expressed
  the dependence on the scattering angle in terms of $t_1$, which is
  in general only possible in the soft limit. In converting to the
  scattering angle, we keep the exact mass dependence in $t_1$
  according to (\ref{eq:tu}), otherwise a $t$-channel singularity
  emerges upon integration over $\theta$.}
\begin{align} 
  \label{eq:massfact} 
  C_{ij}(z,M,m_t,\cos\theta,\mu_f) &= \sum_{a,b}
  C^{ab}_{ij}(z,M,t_1,\mu_f)\otimes \DD_{ab}^{(n_l)}(z,m_t,\mu_f)
  \nonumber
  \\
  &\hspace{1em} \otimes C_{a/t}(z,m_t,\mu_f)\otimes
  C_{b/\bar{t}}(z,m_t,\mu_f)\otimes C_{\fff}(z,m_t,\mu_f) +
  \mathcal{O} \left( \frac{m_t}{M} \right) ,
\end{align} 
where the sum is over all parton species $a,b\in\{t,\bar t,q,\bar
q,g\}$. The functions $C^{ab}_{ij}$ are the partonic cross sections
obtained from the massless scattering process $ij\to ab+\hat{X}$,
where $\hat{X}$ indicates additional final-state partons. The objects
$\DD_{ab}^{(n_l)}$ are defined as the following convolution of
heavy-quark fragmentation functions:
\begin{align}
  \DD_{ab}^{(n_l)}(z,m_t,\mu_f) = D_{a/t}^{(n_l)}(z,m_t,\mu_f)
  \otimes D_{b/\bar{t}}^{(n_l)}(z,m_t,\mu_f) \, .
\end{align}
This convolution of heavy-quark fragmentation functions is completely
analogous to that defining the parton luminosities in
(\ref{eq:lumdef}). It arises after generalizing (\ref{eq:fragdef}) to
a two-fold convolution and performing a change of variables.

We are now ready to discuss the joint limit $z \to 1$ and $m_t/M \to
0$, which is the main theme of this paper. The key point is that these
two limits are independent and commutative, so that we can take them
one-by-one in any order and obtain the same result. We choose to start
from the factorization formula (\ref{eq:massfact}) for the small-mass
limit, and then study the behavior of its component parts in the limit
$z \to 1$. We thus discuss the factorization of the massless
coefficient functions and the fragmentation function in the soft
limit. The alternate method of starting from the factorization formula
(\ref{eq:softfact}) for the soft limit and then studying the
factorization of its component parts in the small-mass limit is
discussed in Appendix~\ref{sec:morefact}.

We first deal with the massless coefficient function $C_{ij}^{ab}$. To
factorize it in the soft limit, we observe that nothing in the
derivation of factorization for the massive coefficient function
(\ref{eq:softfact}) makes reference to the mass of the top-quark.
Therefore, the form of factorization for the massless coefficient
function is exactly the same. The result is thus
\begin{align}
  C_{ij}^{t\bar{t}}(z,M,t_1,\mu_f) = \Tr \left[
    \bm{H}_{ij}(M,t_1,\mu_f) \,
    \bm{S}_{ij}(\sqrt{\hat{s}}(1-z),t_1,\mu_f) \right] +
  \mathcal{O}(1-z) \, .
\end{align}
We have used that only $a=t$ contributes to (\ref{eq:decoupling}) at
leading power in $(1-z)$. The hard function $\bm{H}_{ij}$ is obtained
from virtual corrections to two-to-two scattering with massless top
quarks, and the soft function $\bm{S}_{ij}$ involves only light-like
Wilson lines related to real emission from massless partons. The top
quark is treated as massless in both the external states and in
internal fermion loops, so both the hard and soft function are defined
in a theory with six active massless flavors.

The factorization of the fragmentation functions in the $z \to 1$
limit was explained in \cite{Korchemsky:1992xv,
  Cacciari:2001cw,Gardi:2005yi}, and also within an effective
field-theory framework in \cite{Neubert:2007je}. The main result of
those works is that after the matching onto the $n_l$-flavor theory as
in (\ref{eq:decoupling}), the fragmentation function factorizes into a
product of two functions: one depending on the collinear scale $m_t$,
and the other on the soft-collinear scale $m_t(1-z)$. We write this
factorization as
\begin{align}
  \label{eq:dfact2}
  D_{t/t}^{(n_l)}(z,m_t,\mu_f) = C_D(m_t,\mu_f) \, S_D(m_t(1-z),\mu_f)
  + \mathcal{O}(1-z) \, .
\end{align}
The fragmentation of $\bar{t}$ to $\bar{t}$ follows the same
factorization with the same coefficient functions. The soft function
$S_D$ is related to soft-collinear emission and is equivalent to the
partonic shape-function appearing in $B$-meson decays
\cite{Gardi:2005yi, Neubert:2007je}. The matching coefficient $C_D$ is
independent of $z$ and is a simple function related to virtual
corrections.

Combining all of the information above, the factorization formula for
the partonic cross sections in the joint soft and small-mass limit is
\begin{align}
  \label{eq:fact} 
  C_{ij}(z,M,m_t,\cos\theta,\mu_f) &= C_D^2(m_t,\mu_f) \, \Tr \left[
    \bm{H}_{ij}(M,t_1,\mu_f) \,
    \bm{S}_{ij}(\sqrt{\hat{s}}(1-z),t_1,\mu_f) \right] \nonumber
  \\
  &\hspace{-1em} \otimes C^{ij}_{\fff}(z,m_t,\mu_f) \otimes
  C_{t/t}(z,m_t,\mu_f) \otimes C_{t/t}(z,m_t,\mu_f) \nonumber
  \\
  &\hspace{-1em} \otimes S_D(m_t(1-z),\mu_f) \otimes
  S_D(m_t(1-z),\mu_f) + \mathcal{O}(1-z) + \mathcal{O} \left(
    \frac{m_t}{M} \right) .
\end{align}
The factorization formula (\ref{eq:fact}) is the central result of
this section. In the limit in which it is derived, any choice of
$\mu_f$ generates large logarithms in the soft or small-mass limits.
We deal with this problem in the next section using RG techniques. In
deriving and solving the RG equations it will be useful to introduce
the Laplace-transformed functions
\begin{align}
  \label{eq:LaplaceTransforms}
  \tilde{c}_{ij}(N,M,m_t,\cos\theta,\mu_f) &= \int_0^\infty d\xi \,
  e^{-\xi N} \, C_{ij}(z,M,m_t,\cos\theta,\mu_f) \, , \nonumber
  \\
  \tilde{\bm{s}}_{ij} \Biggl( \ln\frac{M^2}{\bar{N}^2\mu_f^2}, t_1,
  \mu_f \Biggr) &= \int_0^\infty d\xi \, e^{-\xi N} \,
  \bm{S}_{ij}(\sqrt{\hat{s}}(1-z),t_1,\mu_f) \, , \nonumber
  \\
  \tilde{c}_t^{ij} \Biggl( \ln\frac{1}{\bar{N}^2}, m_t, \mu_f \Biggr)
  &= \int_0^\infty d\xi \, e^{-\xi N} \, C^{ij}_{\fff}(z,m_t,\mu_f)
  \nonumber \otimes C_{t/t}(z,m_t,\mu_f) \otimes C_{t/t}(z,m_t,\mu_f)
  \, , \nonumber
  \\
  \tilde{s}_D \Biggl( \ln\frac{m_t}{\bar{N}\mu_f},\mu_f \Biggr) &=
  \int_0^\infty d\xi \, e^{-\xi N} \, S_D(m_t(1-z),\mu_f) \, ,
\end{align}
where $\xi=(1-z)/\sqrt{z}$ and $\bar{N}=Ne^{\gamma_E}$. In Laplace
space, the factorization formula becomes a simple product of the
different functions and reads
\begin{align}
  \label{eq:factL}
  \tilde{c}_{ij}(N,M,m_t,\cos\theta,\mu_f) &= C_D^2(m_t,\mu_f)\, \Tr
  \left[ \bm{H}_{ij}(M,t_1,\mu_f) \, \tilde{\bm{s}}_{ij} \Biggl(
    \ln\frac{M^2}{\bar{N}^2\mu_f^2}, t_1, \mu_f \Biggr) \right]
  \nonumber
  \\
  &\hspace{-2em} \times \tilde{c}_t^{ij} \Biggl(
  \ln\frac{1}{\bar{N}^2}, m_t, \mu_f \Biggr) \, \tilde{s}_D^2 \Biggl(
  \ln\frac{m_t}{\bar{N}\mu_f}, \mu_f \Biggr) + \mathcal{O} \left(
    \frac{1}{N} \right) + \mathcal{O} \left( \frac{m_t}{M} \right) .
\end{align}

\section{The matching functions: fixed-order expansions and RG
  evolution}
\label{sec:resummation}

The component parts of the factorization formula (\ref{eq:fact}) can
be viewed as matching functions in effective theory. In this section
we explain the one-scale calculations needed to extract the
fixed-order expansions of these matching functions, present their RG
equations and the solutions thereof, and give the ingredients needed
to evaluate these RG-improved matching coefficients at NNLL. While
such NNLL calculations require only the NLO perturbative expansion of
the matching coefficients, we also collect all current knowledge at
NNLO, part of which we will use in our numerical analysis later on.

\subsection{Hard function}

The hard function is related to virtual corrections to the two-to-two
scattering processes underlying Born-level production. The method for
calculating the hard-function matrix for the counting $M\sim m_t$ in
fixed-order perturbation theory was described in detail in
\cite{Ahrens:2010zv}, where results valid to NLO were given. This
boiled down to calculating color-decomposed UV-renormalized on-shell
scattering amplitudes and subtracting poles in the $4-d=2\epsilon \to
0$ limit using an IR renormalization factor.

There are in fact two ways to calculate the hard function in the
massless case. The first is to set $m_t=0$ at the start of the
calculation and follow the same procedure as for the massive case.
Then the UV-renormalized scattering amplitudes and the IR
renormalization factors change compared to the massive case, but the
method for extracting the hard function is exactly the same. The
second way is to Taylor-expand the massive result \cite{Ahrens:2010zv}
in the limit $m_t\to 0$ and then convert the result to the massless
case using the relation between massive and massless amplitudes in the
small-mass limit derived in \cite{Mitov:2006xs}. We have calculated
the NLO hard function using both methods and confirmed that they
agree. This NLO result for massless scattering is not new: it is
actually a special case of the more general results for four-parton
scattering given in \cite{Kelley:2010fn}, using the one-loop
calculations of \cite{Ellis:1985er}. We have checked that we reproduce
those results using the procedure described below.

As opposed to the massive case, where only a limited set of NNLO
virtual corrections have been calculated \cite{Czakon:2007ej,
  Czakon:2007wk, Czakon:2008zk, Bonciani:2008az, Bonciani:2009nb,
  Bonciani:2010mn, Korner:2008bn, Kniehl:2008fd}, both the one-loop
times one-loop and two-loop times Born interference terms are known
for the case of massless two-to-two scattering
\cite{Anastasiou:2000kg, Anastasiou:2000mv,Glover:2001af,
  Glover:2001rd}. These results are implicitly summed over colors and
cannot be used to extract the hard matrix directly, on the other hand
they {\it can} be used to extract the contribution of the NNLO hard
function to the approximate NNLO formulas covered below. Although we
do not go through this straightforward but tedious exercise here, it
is an important point that all of the diagrammatic calculations are in
place.

In order to give explicit results valid to NLO we define expansion
coefficients of the hard function $\bm{H}$ as
\begin{align}
  \bm{H} = \alpha_s^2 \frac{3}{8 d_R} \left[ \bm{H}^{(0)}
    +\frac{\alpha_s}{4 \pi} \bm{H}^{(1)} + \left(\frac{\alpha_s}{4
        \pi} \right)^2 \bm{H}^{(2)} +\cdots \right] \, ,
\end{align}
where $d_R = N$ in the quark annihilation channel and $d_R = N^2-1$ in
the gluon fusion channel, with $N=3$ colors in QCD. The LO result
$\bm{H}^{(0)}$ is trivially obtained from the formulas in
\cite{Ahrens:2010zv}, which is regular in the limit $m_t \to 0$. For
the $q\bar q$ channel we have
\begin{align}
  \bm{H}^{(0)}_{q\bar{q}} = \left(
    \begin{matrix}
      0 & 0
      \\
      0 & 2
    \end{matrix}
  \right) \frac{t_1^2+u_1^2}{M^2} \, ,
\end{align}
and the result for the $gg$ channel is
\begin{align}
  \bm{H}^{(0)}_{gg} = \left(
    \begin{matrix}
      \frac{1}{N^2} & \frac{1}{N} \frac{t_1-u_1}{M^2} & \frac{1}{N}
      \\
      \frac{1}{N} \frac{t_1-u_1}{M^2} & \frac{(t_1-u_1)^2}{M^4} &
      \frac{t_1-u_1}{M^2}
      \\
      \frac{1}{N} & \frac{t_1-u_1}{M^2} & 1
    \end{matrix}
  \right) \frac{t_1^2+u_1^2}{2 t_1 u_1} \, .
\end{align}
We do not list the explicit result for the NLO hard function
$\bm{H}^{(1)}$, since it has essentially been given in
\cite{Kelley:2010fn}. For what concerns the quark annihilation
channel, we convert those results to our case by extracting the
elements of the hard matrix from Eq.~(39) of \cite{Kelley:2010fn}.
Subsequently, one needs to consider crossing symmetry and to permute
the arguments of the various matrix elements according to
$H_{ij}(s,t,u) \to H_{ij}(u,s,t)$, as explained in Table~1 of
\cite{Kelley:2010fn}. Finally, one should exchange the element indices
$1 \leftrightarrow 2$ to match the notation employed here, and
compensate for an overall factor, so that
\begin{align}
  \bm{H}^{(0)}_{q\bar{q}} + \frac{\alpha}{4 \pi}
  \bm{H}^{(1)}_{q\bar{q}} = -\frac{1}{64 \pi^2 \alpha_s^2} \left(
    \begin{matrix}
      H_{22}(u,s,t) & H_{21}(u,s,t)
      \\
      H_{12}(u,s,t) & H_{11}(u,s,t)
    \end{matrix}
  \right) \, ,
\end{align}
where all of the elements of the matrix on the right-hand side are
taken from Eq.~(39) of \cite{Kelley:2010fn}. The gluon fusion case is
slightly more complicated because the authors of \cite{Kelley:2010fn}
use a different color basis. In that case, the NLO correction can
obtained from Eqs.~(56) in \cite{Kelley:2010fn} through the rotation
\begin{align}
  \bm{H}^{(1)}_{gg} = 4 O^T   \left( 
    \begin{matrix}
      H_{11}^{NLO} & H_{12}^{NLO} & H_{13}^{NLO}
      \\
      H_{12}^{NLO} & H_{22}^{NLO} & H_{23}^{NLO}
      \\
      H_{13}^{NLO} & H_{23}^{NLO} & H_{33}^{NLO}
    \end{matrix}
  \right) O \, ,
\end{align}
where the matrix elements on the right-hand side are from
\cite{Kelley:2010fn}, and
\begin{align}
  O = \left( 
    \begin{matrix}
      \frac{1}{2 N} & \frac{1}{2} & \frac{1}{2}
      \\
      \frac{1}{2 N} & - \frac{1}{2} &\frac{1}{2}
      \\
      1 & 0 & 0
    \end{matrix}
  \right) \, .
\end{align}

In the resummed formulas below we will need an expression for the hard
function evaluated at an arbitrary scale $\mu_f$, given its value at
an initial scale $\mu_h\sim M$ where it contains no large logarithms
and the fixed-order expansions above can be applied. This is obtained
by deriving and solving the RG equation. As with the matching
coefficient itself, we can derive the RG equation either by setting
$m_t=0$ from the start and using the exact same methods as the massive
case \cite{Ahrens:2010zv}, or we can start from the massive result and
take the limit $m_t \to 0$ using the relations between massive and
massless amplitudes mentioned above. We have checked that the two
methods agree. In any case, the RG equation for the massless hard
function is completely analogous to the massive case studied in
\cite{Ahrens:2010zv} and is given by (here and below we suppress
dependence on the channels $q\bar q$ and $gg$ when there is no
potential for confusion)
\begin{align}
  \label{eq:Hev}
  \frac{d}{d\ln\mu} \bm{H}(M,t_1,\mu) &= \bm{\Gamma}_H(M,t_1,\mu) \,
  \bm{H}(M,t_1,\mu)+ \bm{H}(M,t_1,\mu) \,
  \bm{\Gamma}_H^\dagger(M,t_1,\mu) \, .
\end{align}
The explicit result for the anomalous dimension matrix to two-loop
order in the color basis of \cite{Ahrens:2010zv} is easily derived by
making use of the general result \cite{Becher:2009cu,Becher:2009qa}
for massless scattering amplitudes, and reads
\begin{align}
  \label{eq:qqmatrix}
  \bm{\Gamma}_{q\bar{q}} &= \left[2 C_F \,
    \gamma_{\text{cusp}}(\alpha_s) \, \left( \ln\frac{M^2}{\mu^2} -
      i\pi \right) + 4\gamma^q(\alpha_s) \right] \bm{1} \nonumber
  \\
  & \quad\mbox{} + N \gamma_{\text{cusp}}(\alpha_s) \,
  \left(\ln\frac{-t_1}{M^2}+ i \pi\right)
  \begin{pmatrix}
    0 & 0
    \\
    0 & 1
  \end{pmatrix}
  + \gamma_{\text{cusp}}(\alpha_s) \, \ln\frac{t_1^2}{u_1^2}
  \begin{pmatrix}
    0 & \frac{C_F}{2N}
    \\
    1 & -\frac{1}{N}
  \end{pmatrix}
  ,
\end{align}
and
\begin{align}
  \label{eq:ggmatrix}
  \bm{\Gamma}_{gg} &= \left[ (N+C_F) \, \gamma_{\text{cusp}}(\alpha_s)
    \, \left( \ln\frac{M^2}{\mu^2} -i\pi \right) + 2\gamma^g(\alpha_s)
    + 2\gamma^q(\alpha_s) \right] \bm{1} \nonumber
  \\
  &\quad\mbox{} + N \gamma_{\text{cusp}}(\alpha_s) \, \left(
    \ln\frac{-t_1}{M^2} + i\pi\right)
  \begin{pmatrix}
    0 & 0 & 0
    \\
    0 & 1 & 0
    \\
    0 & 0 & 1
  \end{pmatrix}
  + \gamma_{\text{cusp}}(\alpha_s) \, \ln\frac{t_1^2}{u_1^2}
  \begin{pmatrix}
    0 & \frac{1}{2} & 0
    \\
    1 & -\frac{N}{4} & \frac{N^2-4}{4N}
    \\
    0 & \frac{N}{4} & -\frac{N}{4}
  \end{pmatrix}
  .
\end{align}
For convenience, we have collected results for the various scalar
anomalous dimension functions in Appendix~\ref{app:AnDim}.

Since the evolution equation has the same structure as in the massive
case, it can be solved using the same methods. In presenting the
solution it is convenient to decompose the anomalous dimension into a
logarithmic piece multiplying the unit matrix and a non-logarithmic
part containing the non-trivial matrix structure:
\begin{align}
  \label{eq:gammaH}
  \bm{\Gamma}_H(M,t_1,\mu) = A(\alpha_s) \left( \ln\frac{M^2}{\mu^2} -
    i\pi \right) \bm{1} + \bm{\gamma}^h(M,t_1,\alpha_s) \, ,
\end{align}
where $A=2C_F\gamma_{\text{cusp}}\equiv 2 \Gamma_{\rm cusp}^q$ in the
$q\bar q$ channel and $A=(N+C_F)\gamma_{\rm cusp}\equiv \Gamma_{\rm
  cusp}^g+\Gamma_{\rm cusp}^q$ in the $gg$ channel. The solution to
the RG equation can then be written as
\begin{align}
\label{eq:SolHardRGE}
\bm{H}(M,t_1,\mu) = \bm{U}(M,t_1,\mu_h,\mu) \, \bm{H}(M,t_1,\mu_h) \,
\bm{U}^\dagger(M,t_1,\mu_h,\mu) \, ,
\end{align}
with
\begin{align}
  \label{eq:SolURGE}
  \bm{U}(M,t_1,\mu_h,\mu) &= \exp \bigg[ 2S_A(\mu_h,\mu) -
  a_A(\mu_h,\mu) \left( \ln\frac{M^2}{\mu_h^2} - i\pi \right) \bigg]
  \, \bm{u}(M,t_1,\mu_h,\mu) \, .
\end{align}
The RG exponents are given by
\begin{align}
  \label{eq:Sa}
  S_A(\mu_h,\mu) = -\int\limits_{\alpha_s(\mu_h)}^{\alpha_s(\mu)} \!d\alpha\,
  \frac{A(\alpha)}{\beta(\alpha)} \int\limits_{\alpha_s(\mu_h)}^\alpha
  \!\frac{d\alpha'}{\beta(\alpha')} \, , \qquad a_A(\mu_h,\mu) =
  -\int\limits^{\alpha_s(\mu)}_{\alpha_s(\mu_h)}\!
  d\alpha\,\frac{A(\alpha)}{\beta(\alpha)} \, ,
\end{align}
where $\beta(\alpha_s)=d\alpha_s/d\ln\mu$ is the QCD $\beta$-function
(whose expansion coefficients are given in Appendix~\ref{app:AnDim}).
The matrix-valued contribution to the evolution function reads
\begin{align}
  \label{eq:offdiagu}
  \bm{u}(M,t_1,\mu_h,\mu) = \mathcal{P} \exp
  \int\limits_{\alpha_s(\mu_h)}^{\alpha_s(\mu)}
  \!\frac{d\alpha}{\beta(\alpha)} \, \bm{\gamma}^h(M,t_1,\alpha) \, .
\end{align}
The exact solution to the RG equation is evaluated as a series in
RG-improved perturbation theory as in \cite{Ahrens:2010zv}, where
explicit expressions valid to NNLL order were presented. Such an NNLL
calculation requires the NLO corrections to both the hard matching
function (a one-loop calculation) and the anomalous dimension (a
two-loop calculation).

\subsection{Soft function}

The soft function is related to real emission corrections to massless
$q\bar q,gg\to t\bar t$ scattering in the soft limit. A more formal
definition in terms of the vacuum expectation value of a Wilson-loop
operator was given for the massive case in \cite{Ahrens:2010zv}. The
position-space result for this object can be directly converted into
the Laplace-transformed function (\ref{eq:LaplaceTransforms}). We can
adapt that definition to the massless case simply by changing
time-like Wilson lines representing emissions from massive particles
to light-like Wilson lines representing emissions from massless ones.
We will present results for the Laplace-transformed soft function
directly. We define the perturbative expansion of this function as
\begin{align}
  \label{eq:4piexp}
  \bm{\tilde{s}} = \bm{\tilde{s}}^{(0)} + \frac{\alpha_s}{4 \pi}
  \bm{\tilde{s}}^{(1)} + \left( \frac{\alpha_s}{4 \pi} \right)^2
  \bm{\tilde{s}}^{(2)} + \cdots \,.
\end{align}
At lowest order, the result depends on whether the initial-state
partons are quarks or gluons, but not on the mass. The result is
\begin{align}
  \bm{\tilde{s}}^{(0)}_{q \bar{q}} =
  \begin{pmatrix}
    N & 0
    \\
    0 & \frac{C_F}{2}
  \end{pmatrix}
\end{align}
in the quark annihilation channel, and
\begin{align}
  \bm{\tilde{s}}^{(0)}_{gg} = 
  \begin{pmatrix}
    N & 0& 0
    \\
    0 & \frac{N}{2} & 0
    \\
    0 & 0 & \frac{N^2-4}{2 N}
  \end{pmatrix}
\end{align}
in the gluon fusion channel.

To obtain the Laplace-transformed soft function at NLO we evaluate the
following position-space integrals \cite{Ahrens:2010zv}
\begin{align}
  \label{eq:softintegrals}
  \mathcal{I}_{ij}(\epsilon,x_0,\mu) =
  -\frac{(4\pi\mu^2)^\epsilon}{\pi^{2-\epsilon}} \, v_i \cdot v_j \int
  d^dk \, \frac{e^{-ik^0x_0}}{v_i \cdot k \, v_j \cdot k} \, (2\pi) \,
  \delta(k^2) \, \theta(k^0) \, ,
\end{align}
where $v_i$ are the light-like four-velocities of the partons from the
Born-level scattering process. When $i=j$ these integrals vanish since
$v_i^2=0$, while for $i\neq j$ they are equal to
\begin{align}
  \mathcal{I}_{ij}= -(4 \pi)^{\epsilon} e^{- \epsilon \gamma}
  \Biggl[\frac{2}{\epsilon^2} +\frac{2}{\epsilon} \left(L_0
    -\ln{\frac{v_1 \cdot v_2}{2}} \right) + \left(L_0 -\ln{\frac{v_1
        \cdot v_2}{2}} \right)^2 + \frac{\pi^2}{6} + 2 \Li_2\left( 1-
    \frac{v_1 \cdot v_2}{2}\right) \Biggr]\, ,
\end{align} 
where 
\begin{align}
  L_0 = \ln\left( - \frac{\mu^2 x_0^2 e^{2 \gamma_E}}{4}\right) .
\end{align}
We make use of this result by expressing the scalar products $v_i\cdot
v_j$ in terms of the Mandelstam variables, by subtracting the IR poles
in $\overline{\text{MS}}$, and by Laplace-transforming the integrals
through the replacement $L_0 \to -L$. This leads to
\begin{align}
  \tilde{{\mathcal I}}_{12} &= -\left(L^2 + \frac{\pi^2}{6} \right) ,
  \nonumber
  \\
  \tilde{\mathcal I}_{13} &= \tilde{\mathcal I}_{24} = -\left( L +
    \ln(r) \right)^2 - \frac{\pi^2}{6} - 2\, \Li_2(1-r) \, , \nonumber
  \\
  \tilde{\mathcal I}_{14} &=\tilde{\mathcal I}_{23} = -\left(L +
    \ln(1-r)\right)^2 - \frac{\pi^2}{6} - 2\, \Li_2(r) \, , \nonumber
  \\
  \tilde{\mathcal I}_{11} &=\tilde{\mathcal I}_{22} = \tilde{\mathcal
    I}_{33} = \tilde{\mathcal I}_{44} = 0 \, ,
\end{align}
where $ r = -t_1/M^2$. One obtains the matrix-valued soft function in
Laplace space by evaluating
\begin{align}
  \bm{\tilde{s}}^{(1)} = \sum_{(i,j)} \bm{w}_{ij} \tilde{\mathcal
    I}_{ij} (L,r) \, .
\end{align} 
where the $\bm{w}_{ij}$ are the color matrices from
\cite{Ahrens:2010zv}, which are different for the $q\bar q$ and $gg$
channels, but make no reference to the parton mass.

To obtain the NNLO correction to the soft function requires a new
calculation which is beyond the scope of this paper. However, as a
compromise, we can use the RG equation below to derive all of the
coefficients proportional to powers of logarithms in the
Laplace-transformed NNLO correction, whose form is
\begin{align}
  \bm{\tilde{s}}^{(2)}(L,M,t_1)= \sum_{n=0}^{4}\bm{s}_n(M,t_1) L^n \,
  .
\end{align}
The results for the coefficients are fairly lengthy and since we use
them in this paper only for the factorization check described in
Section~\ref{sec:approxNNLO} we do not list them here.

The Laplace-transformed functions are the central objects used in
solving the RG equations below. One can also convert them to the
momentum-space functions using a set of replacement rules. In the case
where the first argument of the soft function is expressed in terms of
$\sqrt{\hat{s}}(1-z)=M(1-z)/\sqrt{z}$, the resulting distributions are
\begin{align}
  P'_n(z) = \left[\frac{1}{1-z} \ln^n\left(\frac{M^2 (1-z)^2}{\mu^2
        z}\right) \right]_+ .
\end{align}
As shown in \cite{Ahrens:2010zv}, the momentum-space soft function is
derived from the Laplace-space function by making the replacements
\begin{align}
  \label{eq:replacements}
  1 &\to \delta(1-z) \, , \nonumber
  \\
  L &\to 2 P'_0(z) + \delta(1-z) \ln\left(\frac{M^2}{\mu^2} \right) ,
  \nonumber
  \\
  L^2 &\to 4 P'_1(z) + \delta(1-z) \ln^2\left(\frac{M^2}{\mu^2}
  \right) , \nonumber
  \\
  L^3 &\to 6 P'_2(z) - 4 \pi^2 P'_0(z) + \delta(1-z) \left[ \ln^3
    \left(\frac{M^2}{\mu^2} \right) + 4 \zeta_3 \right] , \nonumber
  \\
  L^4 &\to 8 P'_3(z) -16 \pi^2 P'_1(z) + 128 \zeta_3 P'_0(z)
  +\delta(1-z) \left[\ln^4 \left( \frac{M^2}{\mu^2} \right) + 16
    \zeta_3 \ln \left( \frac{M^2}{\mu^2} \right) \right] .
\end{align}
In order to translate the $P'_n$ into the conventional $P_n$
distributions,
\begin{align}
  P_n(z) = \left[\frac{1}{1-z} \ln^n\left(1-z\right) \right]_+ \, ,
\end{align}
we employ the general relation
\begin{align}
  P'_n(z) &= \sum_{k=0}^n \binom{n}{k}
  \ln^{n-k}\left(\frac{M^2}{\mu^2}\right) \Bigg[ 2^k P_k(z)
  \\
  &\quad\mbox{} + \sum_{j=0}^{k-1} \binom{k}{j} 2^j (-1)^{k-j} \left(
    \frac{\ln^j(1-z)\ln^{k-j}z}{1-z} - \delta(1-z) \int_0^1 dx
    \frac{\ln^j(1-x)\ln^{k-j}x}{1-x} \right) \Bigg] \, . \nonumber
\end{align}
The numerical affects of keeping the power-suppressed terms
proportional to $\ln^m z/(1-z)$ was discussed in detail in
\cite{Ahrens:2010zv, Ahrens:2011mw}.

The soft function obeys a non-local RG equation which is solved using
the Laplace transform technique \cite{Becher:2006nr}. The RG
invariance of the total partonic cross section implies that the
Laplace-transformed soft function satisfies
\begin{align}
  \label{eq:Sev}
  \frac{d}{d\ln\mu} \, \tilde{\bm{s}}
  &\left(\ln\frac{M^2}{\mu^2},M,t_1,\mu\right) = \nonumber
  \\
  &- \left[ A(\alpha_s) \ln\frac{M^2}{\mu^2} +
    \bm{\gamma}^{s\dagger}(M,t_1,\alpha_s) \right] \tilde{\bm{s}}
  \left(\ln\frac{M^2}{\mu^2},M,t_1,\mu\right) \nonumber
  \\
  &- \tilde{\bm{s}} \left(\ln\frac{M^2}{\mu^2},M,t_1,\mu\right) \left[
    A(\alpha_s) \ln\frac{M^2}{\mu^2} + \bm{\gamma}^s(M,t_1,\alpha_s)
  \right] .
\end{align}
We have defined
\begin{align}
  \bm{\gamma}^s(M,t_1,\alpha_s) = \bm{\gamma}^h(M,t_1,\alpha_s) +
  \left[2\gamma^{\phi}(\alpha_s) + 2\gamma^{\phi_q}(\alpha_s)\right]
  \bm{1} \, .
\end{align}
Note that while the form of the anomalous dimension is analogous to
that in the massive case, it picks up an extra term
$2\gamma^{\phi_q}$, which is needed to cancel the $\mu$-dependence
from the fragmentation function as determined by the RG equation
(\ref{eq:FragRG}) below.

The solution for the momentum-space soft function reads
\begin{align}
  \bm{S}(\omega,M,t_1,\mu_f) &= \sqrt{\hat s}\, \exp \left[
    -4S_A(\mu_s,\mu_f) + 4a_{\gamma^\phi}(\mu_s,\mu_f)+
    4a_{\gamma^{\phi_q}}(\mu_s,\mu_f) \right] \nonumber
  \\
  &\hspace{-9em} \times \bm{u}^\dagger(M,t_1,\mu_f,\mu_s) \,
  \tilde{\bm{s}}(\partial_{\eta_A},M,t_1,\mu_s) \,
  \bm{u}(M,t_1,\mu_f,\mu_s) \, \frac{1}{\omega}
  \left(\frac{\omega}{\mu_s}\right)^{2\eta_A}
  \frac{e^{-2\gamma_E\eta_A}}{\Gamma(2\eta_A)} \, ,
\label{eq:SolSoftRGE}
\end{align} 
where one is to set $\eta_A= 2 a_A(\mu_s,\mu_f)$ after performing the
derivatives. For values of $2\eta_A<0$, the $\omega$-dependence must
be interpreted in the sense of distributions.

As always in RG-improved perturbation theory, the aim of a formula
such as (\ref{eq:SolSoftRGE}) is to allow one to evaluate the soft
function at an arbitrary scale $\mu_f$ given its result at a scale
$\mu_s$ where it is free of large logarithms. However, the question of
what exactly this $\mu_s$ should be is currently a source of debate in
the literature. We will discuss this issue in more detail when
presenting results for the RG-improved partonic cross section in
Section~\ref{sec:rescs}.

\subsection{Fragmentation function}

The perturbative fragmentation function was calculated at NNLO for
generic values of $z$ in \cite{Melnikov:2004bm}. In this section we
focus on the parts of that result required for the resummed analysis,
namely the leading terms in the soft limit of the function with $n_l$
active flavors defined in (\ref{eq:decoupling}). In particular, we
list results for the functions $S_D$ and $C_D$ appearing in the
factorized form (\ref{eq:dfact2}), determined previously in
\cite{Neubert:2007je}.

The function $S_D$ is related to the partonic shape-function in
$B$-meson decays and can be derived from the two-loop calculations in
\cite{Becher:2005pd}. We define its perturbative expansion in Laplace
space as
\begin{align}
  \label{eq:SDexp}
  \tilde{s}_D = 1 + \frac{\alpha_s}{4 \pi} \tilde{s}_D^{(1)} + \left(
    \frac{\alpha_s}{4 \pi} \right)^2 \tilde{s}_D^{(2)} + \cdots \,.
\end{align}
The expansion coefficients with $N=3$ colors are 
\begin{align}
  \tilde{s}_D^{(1)}(L/2) &= -\frac{4}{3}L^2 -\frac{8}{3} L -
  \frac{10\pi^2}{9} \, ,
  \\
  \tilde{s}_D^{(2)}(L/2) &= \frac{8}{9}L^4 +
  \left(\frac{76}{9}-\frac{8}{27}n_l\right)L^3 +
  \left(-\frac{104}{9}+\frac{76\pi^2}{27}+\frac{16}{27} n_l\right)L^2
  \nonumber
  \\
  & + \left(\frac{440}{27}+\frac{416\pi^2}{27}- 72\zeta_3
    +\frac{16}{81}n_l -\frac{16\pi^2}{27} n_l\right)L \nonumber
  \\
  &-\frac{1304}{81}-\frac{233\pi^2}{9} +\frac{1213\pi^4}{405}
  -\frac{1132\zeta_3}{9}+\left(-\frac{16}{243}+\frac{14\pi^2}{27}
    +\frac{88\zeta_3}{27}\right)n_l \, .
\end{align}
It is a non-trivial check on the factorization formula
(\ref{eq:dfact2}) that the plus-distributions in the fragmentation
function are all related to the momentum-space representation of this
function, obtained by the set of replacement rules analogous to
(\ref{eq:replacements}), but with the substitution $M(1-z)/\sqrt{z}\to
m_t(1-z)$.

The coefficient $C_D$ is related to virtual corrections to the
fragmentation function and is a simple function independent of $z$.
Since the NNLO correction to the fragmentation function obtained in
\cite{Melnikov:2004bm} was not split into real and virtual
corrections, it is not possible to obtain the coefficient directly
from that work. Instead, it must be determined by using the result for
the shape-function given above along with the factorization formula
(\ref{eq:dfact2}) for the fragmentation function. Defining expansion
coefficients in analogy to (\ref{eq:SDexp}), we have (with $N=3$
colors)
\begin{align}
  C_D^{(1)}(m_t,\mu) & = \frac{4}{3}\left(L_m^2 + L_m +
    4+\frac{\pi^2}{6}\right)\,,
  \\
  C_D^{(2)}(m_t,\mu) & = \frac{8}{9}L_m^4 +
  \left(\frac{20}{3}-\frac{8}{27}n_l\right)L_m^3 +
  \left(\frac{406}{9}-\frac{28\pi^2}{27}-\frac{52}{27} n_l\right)L_m^2
  \nonumber
  \\
  & + \left(\frac{2594}{27}+\frac{248\pi^2}{27}- \frac{232\zeta_3}{3}
    -\frac{308}{81}n_l -\frac{16\pi^2}{27} n_l\right)L_m \nonumber
  \\
  &+\frac{21553}{162}+\frac{107\pi^2}{3}-\frac{749\pi^4}{405}
  +\frac{260\zeta_3}{9} +\frac{16\pi^2}{9} \ln 2
  -\left(\frac{1541}{243}+\frac{74\pi^2}{81}
    +\frac{104\zeta_3}{27}\right)n_l \,,
\end{align}
with $L_m = \ln(\mu^2/m_t^2)$. We discuss a possible cross-check of
this result in Appendix~\ref{sec:morefact}.

In the factorization formula for the invariant mass distribution we
need the convolution of two fragmentation functions, which up to NNLO
has the form
\begin{align}
  \DD(z,m_t,\mu)= \delta(1-z) &
  +2\left(\frac{\alpha_s}{4\pi}\right)D^{(1)}(z,m_t,\mu) \nonumber
  \\
  & + \left(\frac{\alpha_s}{4\pi}\right)^2\left[2D^{(2)}(z,m_t,\mu)
    +D^{(1)}(z,m_t,\mu)\otimes D^{(1)}(z,m_t,\mu)\right] .
\end{align}
In order to evaluate the convolutions between the different plus
distributions in the last term we use the formulas derived in
Appendix~\ref{sec:convolutions}.

The RG equation for the fragmentation function is a non-local one. It
is given by
\begin{align}
  \label{eq:FragRG}
  \frac{d}{d\ln\mu}D^{(n_l)}_{t/H}(z,m_t,\mu) = P_{qq}(z,\mu)\otimes
  D^{(n_l)}_{t/H}(z,m_t,\mu)
\end{align}
where $P_{qq}$ is a time-like Altarelli-Parisi splitting function
whose structural form in the soft limit is
\begin{align}
  \label{eq:APev}
  P_{qq}(z,\mu) = \frac{2 \Gamma^q_{\rm cusp}(\alpha_s)}{(1-z)_+}+2
  \gamma^{\phi_q}(\alpha_s)\delta(1-z) \, .
\end{align}
From this equation, and the fact the $S_D$ is equivalent to the
perturbative shape-function from $B$-meson decays \cite{Gardi:2005yi,
  Neubert:2007je}, the RG equations for the function $C_D$ can be
derived. The RG equation for $C_D$ is local, while that for $S_D$ is
non-local and solved using the Laplace transform technique. The result
for the RG-improved fragmentation function reads
\begin{align}
  D(z,m_t,\mu_f) &= \exp\left[2S_{\Gamma_{\rm
        cusp}^q}(\mu_{ds},\mu_{dh})+2a_{\gamma^S}(\mu_{ds},\mu_{dh})
    +2a_{\gamma^{\phi_q}}(\mu_f,\mu_{dh})\right]
  \left(\frac{m_t}{\mu_{ds}}\right)^{-2a_{\Gamma_{\rm
        cusp}^q}(\mu_f,\mu_{dh})} \nonumber
  \\
  &C_D(m_t,\mu_{dh})\tilde{s}_D(\partial_{\eta_d},\mu_{ds})
  \frac{e^{-\gamma_E \eta_d}}{\Gamma(\eta_d)}
  \left(\frac{m_t}{\mu_{ds}}\right)^{\eta_d}\frac{1}{(1-z)^{1-\eta_d}}\,
  ,
\end{align}
with $\eta_d=2a_{\Gamma_{\rm cusp}^q}(\mu_f,\mu_{ds})$. The explicit
results for the anomalous dimension $\gamma^S$ can be found in
\cite{Neubert:2007je} as well as in Appendix~\ref{app:AnDim} of this
paper. While it is obvious that the scale choice $\mu_{dh}\sim m_t$
eliminates large logarithms in the coefficient function $C_D$, the
choice of the scale $\mu_{ds}$ is again a debatable point which we
come back to later on.

\subsection{The heavy-flavor coefficients}

The definition of the heavy-flavor coefficients $C_{t/t}$ and
$C^{ij}_{\fff}$ was given in (\ref{eq:decoupling}). The partonic
matrix elements needed to evaluate those expressions to NNLO are known
from \cite{Melnikov:2004bm} for the fragmentation functions and
\cite{Chuvakin:2001ge} for the PDFs. Rather than give the results
separately, we quote only the result for the Laplace-transformed
combination of the three functions appearing in
(\ref{eq:LaplaceTransforms}). Expressed in terms of $\alpha_s$ with
five active flavors, the result is
\begin{align}
  \tilde{c}_t^{q\bar q}(L,m_t,\mu) & = 1 + n_h \left(
    \frac{\alpha_s}{4\pi} \right)^2 \bigg[ \left( \frac{32}{9} L_m^2 -
    \frac{320}{27} L_m + \frac{896}{81} \right) L \nonumber
  \\
  &\hspace{9em} + \frac{16}{3} L_m^2 - \left( \frac{16}{9} +
    \frac{64\pi^2}{27} \right) L_m + \frac{7592}{243} -
  \frac{64\pi^2}{81} \bigg] + \cdots
  \\
  \tilde{c}^{gg}_{t}(L,m_t,\mu) & = 1 - n_h \frac{\alpha_s}{4\pi}
  \frac{4}{3} L_m + n_h \left( \frac{\alpha_s}{4\pi} \right)^2 \bigg[
  \left( \frac{52}{9} L_m^2 - \frac{520}{27} L_m + \frac{1456}{81}
  \right) L \nonumber
  \\
  & +\frac{8}{3}L_m^2 - \left( \frac{32\pi^2}{27} + \frac{200}{9}
  \right) L_m + \frac{2228}{243} - \frac{16\pi^2}{9}
  +\frac{32\zeta_3}{9} + \frac{4n_h}{9} L_m^2 \bigg] + \cdots
\end{align}
If desired, these can be converted to momentum space by the set of
replacements (\ref{eq:replacements}) with $\mu=M$.

The RG equations for the heavy-flavor coefficients $C_{\fff}$ and
$C_{t/t}$ follow from the fact that the parton luminosity and
fragmentation functions in the $n_h+n_l$ flavor theory obey the
standard Altarelli-Parisi equations in the soft limit. This implies
\begin{align}
  \frac{d}{d\ln\mu}C_{t/t}(z,m_t,\mu) &=
  \left[P^{n_h+n_l}_{qq}(z,\mu)- P^{n_l}_{qq}(z,\mu)\right]\otimes
  C_{t/t}(z,m_t,\mu) \, ,
  \\
  \frac{d}{d\ln\mu} C_{\fff}^{ij}(z,m_t,\mu) &=
  2\left[P^{n_h+n_l}_{ij}(z,\mu)- P^{n_l}_{ij}(z,\mu)\right] \otimes
  C^{ij}_{\fff}(z,m_t,\mu) \, ,
\end{align}
where we have used that non-diagonal evolution is suppressed by powers
of $(1-z)$. The function $P_{gg}$ is defined in analogy with
(\ref{eq:APev}) after the obvious replacements, and both
$\Gamma^g_{\rm cusp}=C_A \gamma_{\rm cusp}$ and $\gamma^{\phi_g}$ can
be read off to two loops from Appendix~\ref{app:AnDim}. The
superscripts indicate the number of active flavors to be used in both
$\alpha_s$ and the coefficients of the anomalous dimensions
themselves. When expressed in terms of a common five-flavor coupling,
the anomalous dimensions $P_{ij}^{n_h+n_l}$ pick up explicit powers of
$\ln m_t/\mu$ related to the $\alpha_s$ decoupling relation
\begin{align}
  \label{eq:asdecoup}
  \alpha_s^{(n_h+n_l)} = \alpha_s^{(n_l)}\left(1 + n_h \frac{2}{3} L_m
    \frac{\alpha_s^{(n_l)}}{4\pi}+\dots \right),
\end{align}
which implies that the anomalous dimension for the heavy-flavor
coefficient no longer has the simple form (\ref{eq:APev}). We note
that for the NNLL analysis, we need only the NLO correction from
$\tilde{c}^{gg}_{t}$, which contains no large logarithms for
$\mu_t\sim m_t$. Beyond NNLL accuracy, however, such a choice leads to
large logarithms in $1-z$ that are not resummed. On the other hand,
since the scale for logarithms related to the decoupling relation
(\ref{eq:asdecoup}) is unambiguously $\mu\sim m_t$, it is conceivable
that one could derive a method for resumming logarithms between the
scales $m_t(1-z)$ and $m_t$ within the heavy-flavor coefficients. We
will however leave this as an open point in our analysis of
resummation in the double soft and small-mass limit.

\section{The partonic cross section at NNLL and approximate NNLO}
\label{sec:rescs}

In this section we derive the final expression for the resummed
partonic cross section at NNLL in the double soft and small-mass
limit. We discuss a few points having to do with its practical
implementation, and then turn to its approximate NNLO implementation.

\subsection{Partonic cross section at NNLL}

To derive the final result for the resummed partonic cross section at
a scale $\mu_f$ we insert the RG-improved results for the hard, soft,
and fragmentation functions presented above into the factorization
formula (\ref{eq:fact}). The convolution integrals can be performed
analytically using
\begin{align}
  \int_{z}^{1} dz'
  \frac{1}{(1-z')^{1-\eta_1}}\frac{1}{(1-z/z')^{1-\eta_2}} \approx
  \frac{\Gamma(\eta_1)\Gamma(\eta_2)}{\Gamma(\eta_1+\eta_2)}\frac{1}{(1-z)^{1-\eta_1-\eta_2}}
  \, ,
\end{align}
where the approximation is true in the limit $z\to 1$. We can also
simplify the various products of evolution matrices by employing the
relations
\begin{align}
  \bm{u}\left(M, \cos \theta, \mu_f,\mu_s \right) \bm{u}\left(M, \cos
    \theta,\mu_h,\mu_f \right) &= \bm{u}\left(M, \cos
    \theta,\mu_h,\mu_s \right) \, , \nonumber
  \\
  a_A(\mu_s,\mu_h) +a_A(\mu_h,\mu_f) & = a_A (\mu_s,\mu_f) \, ,
  \nonumber
  \\
  S_A\left(\mu_h,\mu_f\right) -S_A\left(\mu_s,\mu_f\right) &=
  S_A\left(\mu_h,\mu_s \right) -a_A \left(\mu_s,\mu_f \right)
  \ln{\frac{\mu_h}{\mu_s}} \, .
\end{align}

The only subtlety is related to the treatment of heavy-quark threshold
effects at $\mu_t\sim m_t$. However, this is a standard problem in
RG-improved perturbation theory involving heavy quarks, and we deal
with it in the usual way \cite{Buchalla:1995vs}. To understand the
logic, it suffices to consider a hypothetical observable involving
three widely separated scales $M\gg m_t \gg \mu_0$, which satisfies a
factorization formula of the form $C(M,\mu)D(m_t,\mu) F(\mu_0,\mu)$.
If $C$ is a coefficient function whose RG running is known, and the
goal is to evolve it from a high scale $\mu_M\sim M$ to the scale
$\mu_0$ below the heavy-flavor threshold, one uses the following
schematic equation
\begin{align}
  \label{eq:splitrunning}
  C^{(n_l)}(M,m_t,\mu_0)=
  U^{(n_l)}(M,\mu_0,\mu_t)M_h(m_t,\mu_t)U^{(n_l+n_h)}
  (M,\mu_t,\mu_M)C^{(n_l+n_h)}(M,\mu_M) \, .
\end{align}
In words, one uses six-flavor evolution functions $U^{(n_l+n_h)}$
above the flavor threshold at $\mu_t$, and five-flavor evolution
functions $U^{(n_l)}$ below it. The change in the number of flavors
induces a matching coefficient $M_h$ at the flavor threshold. It is
determined by requiring that the partonic cross section in the $n_f$
and $n_f-1$ flavor theories be equal at the scale $\mu_t$:
\begin{align}
  C^{(n_f)}(M,\mu_t) \langle D F \rangle^{(n_f)} =
  C^{(n_f-1)}(M,m_t,\mu_t) \langle D F \rangle^{(n_f-1)} \, ,
\end{align}
where the $\langle \rangle^{(f)}$ denotes a partonic matrix element in
the theory with $f$ massless flavors evaluated at the scale $\mu_t$.

The generalization of this simple picture to our case is
straightforward. The role of the coefficient $C$ is played by the
massless hard and soft functions, and that of the heavy-flavor
matching coefficient $M_h$ by the convolution of the three functions
appearing in the second line of (\ref{eq:fact}). The explicit result
reads
\begin{align}
  \label{eq:MasterFormula}
  C(z,M,m_t, \cos\theta, \mu_f) &=\exp\bigg[4 S_{\Gamma_{\rm
      cusp}^q}(\mu_{ds},\mu_{dh}) +4 a_{\gamma^{\phi}}(\mu_t,\mu_f)+ 4
  a_{\gamma^{\phi_q}}(\mu_t,\mu_{dh}) +4
  a_{\gamma^S}(\mu_{ds},\mu_{dh})\nonumber
  \\
  & +2 a_{\Gamma_{\rm
      cusp}^q}(\mu_{dh},\mu_{ds})\ln\frac{m_t^2}{\mu_{ds}^2}\bigg]_{n_f=5}\,
  \exp[4 a_{\gamma^{\phi}}(\mu_s,\mu_t)+ 4
  a_{\gamma^{\phi_q}}(\mu_s,\mu_{t})]_{n_f=6} \nonumber
  \\
  & \times \Tr \Bigg[ \bm{U}(M,t_1,\mu_h,\mu_s) \, \bm{H}(M,t_1,\mu_h)
  \, \bm{U}^\dagger(M,t_1,\mu_h,\mu_s) \nonumber
  \\
  & \times \tilde{\bm{s}}
  \left(\ln\frac{M^2}{\mu_s^2}+\partial_{\eta'},M,t_1,\mu_s\right)
  \Bigg]_{n_f=6} \nonumber
  \\
  & \times \left[ \tilde{c}_t^{ij}(\partial_{\eta'},m_t,\mu_t)
    \widetilde{\dd}
    \left(\ln\frac{m_t^2}{\mu_{ds}^2}+\partial_{\eta'},m_t,\mu_{dh},\mu_{ds}\right)
  \right]_{n_f=5} \frac{e^{-2\gamma_E \eta'}}{\Gamma(2\eta')}
  \frac{1}{(1-z)^{1-2\eta'}} \nonumber
  \\
  & + {\cal O}(1-z)+{\cal O}\left(\frac{m_t}{M}\right)\, .
\end{align}
We have defined $\eta'=\left[2 a_A(\mu_s,\mu_t)\right]_{n_f=6}
+\left[2a_A(\mu_t,\mu_f) + 2a_{\Gamma_{\rm
      cusp}^q}(\mu_f,\mu_{ds})\right]_{n_f=5}$ and in addition
$\widetilde{\dd}(L,m_t,\mu_{dh},\mu_{ds})=
\left[C_D(m_t,\mu_{dh})\tilde{s}_D(L/2,\mu_{ds})\right]^2 $. We have
indicated with the subscripts the number of active massless flavors
$n_f$ to be used in evaluating the running coupling constant and
perturbative functions in the various parts of the formula. This
number of active flavors is chosen according to the physical picture
of the schematic example (\ref{eq:splitrunning}) above, namely that of
integrating out heavy degrees of freedom until reaching a scale under
the flavor threshold below which the remaining degrees of freedom are
factorized into the PDFs. However, it is formally true for any value
of the factorization scale. It thus provides a convenient way to use
the standard PDFs with five light flavors even when $\mu_f$ is far
above the heavy-flavor threshold at $\mu_t\sim m_t$, as it explicitly
resums any large logs in $m_t/M$ in the partonic cross section for
such a scale choice.

The result (\ref{eq:MasterFormula}) is the final expression for the
resummed partonic cross section in momentum space. It can be evaluated
perturbatively at NNLL order using the results given in the previous
section. There are two important issues in terms of its numerical
evaluation. The first is a technical one having to do with the choice
of matching scales, the second is a practical one having to do with
the power corrections away from the soft and small-mass limits. We end
this section on the resummed cross section by discussing these in
turn.

As alluded to several times in the previous section, the philosophy of
RG-improved perturbation theory is to use RG evolution factors to
evaluate the matching functions at an arbitrary scale $\mu_f$ given
their value at an initial scale where they do not involve large
logarithms. This RG running then exponentiates large corrections
appearing when $\mu_f$ is parametrically far from the natural scale.
For the hard function and the coefficient $C_D$, which obey local RG
equations, the choice is straightforward: one uses $\mu_h\sim M$ and
$\mu_{dh}\sim m_t$. For the massless soft function and the function
$S_D$, the correct choice of this scale is less obvious. If the goal
is to resum logarithms of $(1-z)$ in the partonic cross sections, then
the natural scales are $M(1-z)$ and $m_t(1-z)$. However, such choices
are ill-defined at the level of the momentum-space result
(\ref{eq:MasterFormula}). Partonic logarithms can be resummed by
choosing the scales $\mu_s$ and $\mu_{ds}$ at the level of the
Laplace-transformed functions (\ref{eq:LaplaceTransforms}) and
performing the inverse transform back to momentum space numerically,
but at the cost of introducing the Landau-pole singularity familiar
from Mellin-space implementations of soft-gluon resummation for
top-quark pair production \cite{Laenen:1991af, Laenen:1993xr,
  Berger:1995xz, Berger:1996ad, Berger:1997gz, Catani:1996dj,
  Bonciani:1998vc}. An alternate method is to choose the two scales as
numerical functions of $M$, in such a way that the logarithmic
corrections to the hadronic cross section arising from those in the
partonic one are minimized after convolutions with the PDFs
\cite{Becher:2007ty}. The fixed-order expansions of the resulting
expressions at any finite order in the logarithmic counting are then
of a different structure than those in the partonic cross section
\cite{Ahrens:2011mw, Bonvini:2012yg}. Studying the numerical
differences between these methods would be an interesting exercise,
but since we will not do detailed phenomenology in the current paper
we leave this issue aside.

Dealing with the power corrections away from the double soft and
small-mass limit is also important, although considerably more
straightforward technically. The standard method is to include these
corrections at NLO in fixed-order perturbation theory, thus obtaining
NLO+NNLL accuracy. This is accomplished by evaluating partonic cross
sections $d\hat\sigma$ as
\begin{align}
  d\hat\sigma(\mu_f) = d\hat\sigma^{\rm NNLL}(\{\mu_i\})\bigg|_{m_t
    \to 0,z\to 1}+ \left(d\hat\sigma^{\rm NLO}(\mu_f)
    -d\hat\sigma^{\rm NNLL}(\{\mu_i\}=\mu_f\})\bigg|_{m_t \to 0,z\to
      1}\right)\,,
\end{align}
where by $\{\mu_i\}$ we mean the set of scales $\mu_h,\mu_s,\dots$
appearing in (\ref{eq:MasterFormula}). Such a resummation formula is
useful only at values of the invariant mass where $m_t/M$ is truly
small. It is straightforward to extend the matching procedure to take
into account in addition the set of higher-order $m_t/M$ corrections
determined from soft gluon resummation at NNLL with the counting
$m_t\sim M$ used in \cite{Ahrens:2010zv}, thus yielding a result
useful for the full range of $M$, but we do not give the explicit
results here.

\subsection{Partonic cross section at approximate NNLO}
\label{sec:approxNNLO}

The resummed formula from the previous subsection can be used as a
means of extracting what can be argued to be dominant part of the full
NNLO correction in fixed-order perturbation theory. This truncation of
the resummed expansion is of course not valid if the logarithms are
truly large, but it is easy to imagine a situation where $(1-z)$ and
$m_t/M$ are good expansion parameters for the fixed-order corrections,
but not necessary so small that logarithmic corrections beyond NNLO
are numerically important. In this section we focus on such
approximate NNLO formulas based on our NNLL results, and explain how
the results in this paper can offer an improvement on those previously
derived in \cite{Ahrens:2009uz, Ahrens:2010zv}.

It is convenient to discuss the approximate NNLO formulas at the level
of Laplace-transformed coefficients. The general expression for the
NNLO correction in Laplace space reads
\begin{align}
  \label{eq:C2L}
  \tilde{c}(N,M,m_t,\cos\theta,\mu_f) &=
  \alpha_s^2\bigg[\tilde{c}^{(0)}(M,m_t,\cos\theta,\mu_f)
  +\left(\frac{\alpha_s}{4 \pi} \right)
  \tilde{c}^{(1)}(N,M,m_t,\cos\theta,\mu_f) \nonumber
  \\
  &+ \left(\frac{\alpha_s}{4 \pi} \right)^2
  \tilde{c}^{(2)}(N,M,m_t,\cos\theta,\mu_f) + {\mathcal O}(\alpha_s^3)
  \bigg] \, , \nonumber
  \\
  \tilde{c}^{(2)}(N,M,m_t,\cos\theta,\mu_f) &= \sum_{n=0}^4
  c^{(2,n)}(M,m_t,t_1,\mu_f) \, \ln^n\frac{M^2}{\bar{N}^2\mu_f^2} +
  \mathcal{O}\left(\frac{1}{N}\right) .
\end{align}
The terms $c^{(2,n)}$ for $n=1,2,3,4$ are determined exactly by NNLL
soft-gluon resummation for arbitrary $m_t$ \cite{Ahrens:2009uz}. On
the other hand, only parts of the $c^{(2,0)}$ coefficient are
determined by the NNLL calculation, namely its $\mu$-dependence and
the contribution from the product of NLO corrections to the hard and
soft functions. To determine this coefficient exactly would require
the massive hard and soft functions at NNLO in fixed-order
perturbation theory, which are parts of the N$^3$LL calculation.

We now discuss to what extent the results from this paper can improve
the NNLO approximation described above. For the pieces proportional to
the Laplace-space logarithms there is no possible improvement to the
exact results as a function of $m_t$. However, it is a non-trivial
check on the factorization formula (\ref{eq:factL}) that its expansion
in fixed-order reproduces these results in the limit $m_t/M\to 0$. We
have confirmed explicitly that this is the case, also for the $n_h$
pieces after converting the results for the massless hard and soft
functions to a theory with five active flavors using
(\ref{eq:asdecoup}). For the $c^{(2,0)}$ term, on the other hand, we
can take the further step of determining exactly the terms enhanced by
logarithms of $M/m_t$. Moreover, since the NNLO fragmentation function
is known, and all diagrammatic calculations needed to extract the
contribution of the two-loop hard function to this term are in place,
the only piece needed to fully determine this coefficient in the limit
$m_t/M\to 0$ is the NNLO massless soft function. This is a much
simpler calculation than that for generic top-quark mass, and once
completed it will provide valuable insight into the uncertainties in
approximate NNLO calculations based on NNLL resummation alone.

Given results for the Laplace-space coefficients, we can obtain a
result in momentum space by making the replacements in
(\ref{eq:replacements}). The momentum-space coefficient then takes the
form
\begin{align}
  \label{eq:C2}
  C^{(2)}(z,M,m_t,\cos\theta,\mu) &= D_3 \left[\frac{\ln^3 (1-z)}{1-z}
  \right]_+ + D_2 \left[\frac{\ln^2(1-z)}{1-z} \right]_+ \nonumber
  \\
  &\quad + D_1 \left[\frac{\ln(1-z)}{1-z}\right]_+ + D_0
  \left[\frac{1}{1-z} \right]_+ + C_0\,\delta(1-z) + R(z) \, .
\end{align}
The coefficients $D_0,\ldots, D_3$ and $C_0$ are functions of the
variables $M,m_t,t_1,$ and $\mu$. The plus distributions $D_i$ are
determined by the approximate NNLO formula in Laplace space; explicit
results were given in \cite{Ahrens:2009uz}. These plus-distribution
coefficients are exact, valid for generic values of the top-quark
mass. To determine the delta-function coefficient exactly would also
require the unknown pieces of the $c^{(2,0)}$ coefficient, in other
words the NNLO hard and soft functions. Without these pieces, there is
an ambiguity as to what to include in this term. Parts of these are
directly related to those in Laplace transformed coefficient
$c^{(2,0)}$, and parts are related to whether to include the results
from inverting the Laplace transform via the replacements
(\ref{eq:replacements}); we will always specify our means of dealing
with these ambiguities when giving numerical results later on.

The decomposition (\ref{eq:C2}) is the natural one for studying the
NNLO correction within the framework of soft-gluon resummation.
However, it is worth mentioning that another way to estimate
higher-order corrections would be to instead focus on the leading
terms in the $m_t/M$ expansion, as determined by factorization formula
(\ref{eq:massfact}) for the small-mass limit. Given NNLO fragmentation
functions and the NLO massless scattering kernels $C_{ij}$ for {\it
  generic} values of $z$, it would then be possible to study to what
extent the leading terms in the soft limit reproduce the singular
logarithmic corrections in the limit $m_t/M\to 0$. While such a
calculation would be interesting and valuable as a means of studying
power corrections to soft-gluon resummation, it is clearly beyond the
scope of this work to pursue this idea further.

\section{Numerical studies}
\label{sec:numerics}

In this section we perform short numerical studies of our results.
Since a detailed analysis of soft-gluon resummation for the
pair-invariant mass distribution was carried out with the counting
$m_t\sim M$ in \cite{Ahrens:2009uz, Ahrens:2010zv}, our main
motivation is to study to what extent these results can be improved
through the additional layer of resummation in the small-mass limit
$m_t\ll M$.

When arguing that resummation is required in a certain limit, a
typical first step is to check to what extent the corrections at a
given order in perturbation theory are related to the logarithmic
pieces. If the logarithms account for the bulk of the corrections, it
is an obvious improvement to use the resummed formula to include
subsets of higher-order corrections related to them. Comparisons of
NLO corrections in the $z\to 1$ limit with exact fixed-order results
were performed in \cite{Ahrens:2009uz, Ahrens:2010zv, Ahrens:2011mw}.
It was shown there that the logarithmic plus distributions, determined
by an NLL calculation, account for a bit more than half of the NLO
corrections, while these logarithms plus the delta-function term,
determined by an NNLL calculation, account for essentially all of it.
Moreover, it was observed that the perturbative corrections at the
scale $\mu_f=M$ are rather large at high values of the invariant mass.
The main question we seek to answer in this section is whether this is
due to the small-mass logarithms. If so, it would be necessary to
supplement the phenomenological results of \cite{Ahrens:2010zv} with
the small-mass resummation derived here.

We address this question at NLO by isolating the terms which can give
rise to large logarithms in $m_t/M$. This is easily done by expanding
the resummed formula (\ref{eq:MasterFormula}) to NLO in
$\alpha_s(\mu_f)$, for the choice $\mu_h=\mu_s = M$,
$\mu_{dh}=\mu_{ds}=\mu_{t}=m_t$. The NLO corrections proportional to
mass logarithms are determined by the NLL calculation, and when
expressed in a theory with five active massive flavors, they read
(normalized to the Born-level Laplace-space coefficient $\tilde
c^{(0)}$)
\begin{align}
  \label{eq:NNLexp}
  & \frac{2\alpha_s}{4\pi}\bigg\{ \left( \Gamma_{{\rm
        cusp},0}^q\ln\frac{M^2}{m_t^2} +\Gamma_{{\rm
        cusp},0}\ln\frac{M^2}{\mu_f^2} \right) \left[\frac{1}{1-z}
  \right]_+ \nonumber
  \\
  &+ \bigg( \left(\frac{2}{3}\delta_{q\bar q}N_h +
    \gamma^{\phi_q}_{0}\right)\ln\frac{M^2}{m_t^2} +\left(-\beta_0 +
    \gamma_0^\phi\right)\ln\frac{M^2}{\mu_f^2}\bigg)\delta(1-z)\bigg\}
  \,.
\end{align}
We have checked numerically that for $\mu_f\sim M$ these small-mass
logarithms make up only a small part of the NLO corrections in the
$z\to 1$ limit (which in turn account for most of the full
correction), even for values of $M$ as high as 3$-$5~TeV at the LHC
with $\sqrt{s}=7$~TeV. The NLL corrections determined by soft-gluon
resummation for arbitrary $m_t$, which include also plus distributions
containing no small-mass logarithms, make up a much larger part of the
exact NLO correction. We will therefore view small-mass resummation as
supplementary to that, a means of improvement rather than a
substitute. The point is that the mass logarithms generated at NLL in
the small-mass limit are a subset of the NNLL corrections in soft
gluon resummation for generic $m_t$, so including them may be
advantageous. We now study this statement in more detail.

To do so, we define different approximations to the NLO corrections at
the level of the Laplace-space coefficient, which reads
\begin{align}
  \tilde{c}^{(1)}(N,M,m_t,\cos\theta,\mu_f) &= \sum_{n=0}^2
  c^{(1,n)}(M,m_t,t_1,\mu_f) \, \ln^n\frac{M^2}{\bar{N}^2\mu_f^2} +
  \mathcal{O}\left(\frac{1}{N}\right) .
\end{align}
Momentum space results are obtained through the replacement rules
(\ref{eq:replacements}). The coefficients $c^{(1,2)}$ and $c^{(1,1)}$
thus determine the plus-distribution coefficients. We will take the
exact expression for these, valid for arbitrary $m_t$. We then
distinguish between three approximations, which differ only in their
treatment of $c^{(1,0)}$:
\begin{enumerate}
\item use no information, i.e.  $c^{(1,0)}=0$;
\item use the information from the NLO fragmentation function, the
  heavy-flavor matching coefficients, and $\alpha_s$ decoupling in the
  massless hard and soft functions, thereby including all terms
  enhanced by $\ln m_t/M$ for $\mu_f \sim M$;
\item use the information from approximation 2 plus the constant pieces of 
the massless NLO hard and soft functions.
\end{enumerate}
These three approximations add progressively more information to the
Laplace-space coefficients: the first is NLL in soft gluon
resummation, the second contains parts of the NNLL correction enhanced
by small-mass logarithms, and the third includes the entire NNLL
correction but expanded in the limit $m_t/M\to 0$. We show the NLO
correction to the invariant mass distribution obtained from these
three approximations in Table~\ref{tab:nloapprox}. We also show the
``exact'' result obtained from the leading terms in the $z\to 1$
limit, for arbitrary top-quark mass.
\begin{table}[h!]
  \centering
  \begin{tabular}{|c|c|c|c|}
    \hline
    & $M=500$~GeV & $M=1500$~GeV & $M=3000$~GeV
    \\ \hline
    Approx. 1 & 0.074 & $1.04 \times 10^{-4}$ & $0.70 \times 10^{-7}$
    \\ \hline
    Approx. 2 & 0.085 & $1.35 \times 10^{-4}$ & $0.94 \times 10^{-7}$
    \\ \hline
    Approx. 3 & 0.126 & $1.79 \times 10^{-4}$ & $1.19 \times 10^{-7}$
    \\ \hline
    Exact ($z\to 1$) & 0.154 & $1.86 \times 10^{-4}$ & $1.20 \times 10^{-7}$
    \\ \hline
  \end{tabular}
  \caption{NLO corrections to the differential cross section
    $d\sigma/dM$ (in pb/GeV) computed using $m_t=172.5$~GeV, $\mu_f=M$
    and MSTW2008NNLO PDFs, at LHC with $\sqrt{s}=7$~TeV. }
  \label{tab:nloapprox}
\end{table}
We see that the first approximation accounts for about 50-60\% of the
exact answer. The second approximation is a small improvement, and can
account for up to 75\% of the exact answer for high invariant mass.
The third is a big improvement over the first two, especially at lower
values of the invariant mass, and accounts for nearly the entire
correction in the $z\to 1$ limit already at $M$=1500~GeV.

We now discuss approximate NNLO corrections, and how they change upon
including extra information from the small-mass limit. Following our
NLO analysis, we take the Laplace-transformed coefficient
(\ref{eq:C2L}) as the fundamental object. As mentioned in the previous
section, the terms $c^{(2,n)}$ for $n=1,2,3,4$ are known exactly for
arbitrary $m_t$, since they are part of the NNLL calculation. We keep
the full mass dependence of these terms in our NNLO approximation, and
convert them to momentum-space results as in (\ref{eq:replacements}).
We then consider two approximations for the constant term $c^{(2,0)}$:
\begin{enumerate}
\item[A.] use no information, i.e. $c^{(2,0)}=0$;
\item[B.] use the information from the NNLO fragmentation function,
  plus the $n_h$ terms arising from $\alpha_s$-decoupling and the
  heavy-flavor coefficients, thereby including all terms enhanced by
  (up to two) powers of $\ln m_t/M$ for $\mu_f\sim M$.
\end{enumerate}
The first of these is a pure NNLL calculation in soft gluon
resummation for arbitrary $m_t$, while the second is NNLL plus the
part of the N$^3$LL correction enhanced by small-mass logarithms (as
well as some unenhanced terms associated with the fragmentation
function, heavy-flavor coefficients, and $\alpha_s$-decoupling to
second order). The results for the NNLO correction corresponding to
these choices are shown in Table~\ref{tab:nnloapprox}.
\begin{table}[t]
  \centering
  \begin{tabular}{|c|c|c|c|}
    \hline
    & $M=500$~GeV & $M=1500$~GeV & $M=3000$~GeV
    \\ \hline
    Approx. A & $5.67 \times 10^{-2}$ & $1.22 \times 10^{-4}$ & $ 1.11\times 10^{-7}$
    \\ \hline
    Approx. B & $6.35 \times 10^{-2}$ & $1.40 \times 10^{-4}$ & $ 1.26 \times 10^{-7}$
    \\ \hline
  \end{tabular}
  \caption{\label{tab:nnloapprox} NNLO corrections to the differential cross section
    $d\sigma/dM$ (in pb/GeV) computed using $m_t=172.5$~GeV, $\mu_f=M$
    and MSTW2008NNLO PDFs, at LHC with $\sqrt{s}=7$~TeV.}
\end{table}
The two approximations are numerically rather close to one another.
This shows that logarithms enhanced by powers of $\ln m_t/M$ do not
lead to large corrections. Moreover, at higher values of the invariant
mass the NNLO corrections can be even larger than the NLO ones. This
motivates all-orders soft-gluon resummation instead of NNLO
expansions, which was indeed the approach taken in
\cite{Ahrens:2010zv}. However, the only way to know the size of the
missing N$^3$LL corrections is to calculate them, and it will be
interesting to return to this issue once they are known in the
$m_t/M\to 0$ limit.

\section{Conclusions}
\label{sec:conclusions}

The pair invariant-mass distribution is an important observable for
top-quark physics at hadron colliders. In this paper we set up a
framework for dealing with potentially large perturbative corrections
at high values of invariant mass. In particular, we gave explicit
factorization and resummation formulas appropriate in the double soft
($z\to 1$) and small-mass ($m_t\ll M$) limit of the differential cross
section, along with the ingredients needed to evaluate them to NNLL
order. While many of the ideas and perturbative calculations needed to
accomplish such a resummation were already available in the
literature, this is the first time they have all been combined for a
description of the invariant-mass distribution in top-quark pair
production at hadron colliders. With small modifications the methods
can also be used for the double soft and small-mass limit of
single-particle inclusive observables such as the $p_T$ or rapidity
distribution of the top quark.

We deferred a detailed phenomenological study of our results to future
work. However, a short numerical study revealed the following
features. First, even for invariant mass values as high as 3~TeV, it
is not obvious that small-mass logarithms of the ratio $m_t/M$ are so
large that they need to be resummed. On the other hand, already for
values of the invariant mass of around 1.5~TeV, the leading terms in
the $m_t/M\to 0$ limit, including the constant pieces as well as the
logarithms, provide an excellent approximation to the full correction
within the soft limit. We thus envision the main utility of the
results obtained here as a means of adding parts of the N$^3$LL
corrections to soft-gluon resummation for arbitrary values of $m_t$,
as an expansion in $m_t/M$. The missing piece of this analysis is the
NNLO soft function for massless two-to-two scattering. Once completed,
the calculation of this function will provide valuable information
into the importance of higher-order corrections to the NLO+NNLL
results for the invariant mass distribution presented in
\cite{Ahrens:2010zv}.

\section*{Acknowledgments}

We would like to thank Matthias Neubert for useful discussions. This
research was supported in part by the PSC-CUNY Award N. 64133-00 42,
by the NSF grant PHY-1068317, by the German Research Foundation under
grant NE398/3-1, 577401: {\em Applications of Effective Field Theories
  to Collider Physics}, by the European Commission through the {\em
  LHCPhenoNet\/} Initial Training Network PITN-GA-2010-264564, and by
the Schweizer Nationalfonds under grant 200020-124773.

\newpage

\appendix

\section{Small-mass limits of massive hard and soft functions}
\label{sec:morefact}

Our derivation of the factorization formula (\ref{eq:fact}) used as a
starting point the small-mass factorization formula
(\ref{eq:massfact}) for the differential partonic cross section. In
this appendix we briefly discuss the alternative derivation starting
from the factorization formula (\ref{eq:softfact}) and studying the
properties of the massive hard and soft functions in the small-mass
limit. For simplicity, we neglect contributions of heavy-quark loops,
which give rise to additional terms proportional to $n_h$ taken into
account by the heavy-flavor matching coefficient
(\ref{eq:LaplaceTransforms})

We begin by recalling that the massive hard matrix is related to
UV-renormalized virtual corrections to color-decomposed amplitudes for
two-to-two scattering. These IR-divergent quantities are rendered
finite through multiplication by a renormalization matrix $\bm{Z}_m$.
We define the bare and renormalized quantities according to
\begin{align}
  \label{eq:Zm}
  \lim_{\epsilon \to 0} \bm{Z}_m^{-1}(\epsilon,M,m_t,t_1,\mu)
  \Ket{\mathcal{M}(\epsilon,M,m_t,t_1)} =
  \Ket{\mathcal{M}_{\text{ren}}(M,m_t,t_1,\mu)},
\end{align}
where the quantity on the right is finite in the limit $\epsilon\to
0$. We write the analogous relation for massless amplitudes as
\begin{align}
  \label{eq:Z0}
  \lim_{\epsilon \to 0} \bm{Z}^{-1}(\epsilon,M,t_1,\mu)
  \Ket{\mathcal{M}(\epsilon,M,t_1)} =
  \Ket{\mathcal{M}_{\text{ren}}(M,t_1,\mu)} \,.
\end{align}
We can use these together with the relation between massive and
massless scattering amplitudes in the small-mass limit
\cite{Mitov:2006xs} to derive a relation between the massless and
massive hard functions. To do so, we first use
\begin{align}
  \label{eq:mmfact}
  \Ket{\mathcal{M}(\epsilon,M,m_t,t_1)}=
  Z_{[q]}(\epsilon,m_t,\mu)\Ket{\mathcal{M}(\epsilon,M,t_1)}
\end{align}
where $Z_{[q]}$ is the object given to NNLO in Eqs. (37) and (38) of
\cite{Mitov:2006xs}. Here and in the remainder of the section we have
neglected terms which vanish in the limit $m_t\to 0$. We then multiply
both sides of (\ref{eq:mmfact}) by the massive renormalization factor,
use (\ref{eq:Zm}) to deduce that both sides are finite, and then
observe that (\ref{eq:Z0}) implies that
\begin{align}
  \label{eq:Zmatching}
  Z_{[q]}(\epsilon,m_t,\mu) \bm{Z}_m^{-1}(\epsilon,M,m_t,t_1,\mu) &=
  f(m_t,\mu)\bm{Z}^{-1}(\epsilon,M,t_1,\mu) \,.
\end{align}
The function $f$ is a scalar matching correction which is finite in
the limit $\epsilon\to 0$. From this relation and the definition of
the hard matrix in terms of the IR-renormalized color decomposed
amplitudes, it then follows that
\begin{align}
  \bm{H}_{ij}^m(M,m_t,t_1,\mu) = f^2(m_t,\mu) \bm{H}_{ij}(M,t_1,\mu)
  \,.
\end{align}
Since when neglecting terms proportional to $n_h$ the only $m_t$
dependence in the factorization formula for the partonic cross section
is through the fragmentation function, it must also be true that
\begin{align}
  \label{eq:softfactmassless}
  \bm{S}_{ij}^m (\sqrt{\hat{s}}(&1-z),m_t,t_1,\mu) =
  \\
  &\bm{S}_{ij}(\sqrt{\hat{s}}(1-z),t_1,\mu) \otimes
  \frac{C_D(m_t,\mu)S_D(m_t(1-z),\mu)}{f(m_t,\mu)}\otimes
  \frac{C_D(m_t,\mu)S_D(m_t(1-z),\mu)}{f(m_t,\mu)} \,. \nonumber
\end{align}
Finally, we recall that the soft function is related to real gluon
emission in the soft limit, a statement independent of whether the
final-state quarks are massive or massless. Since all real emission
contributions in the fragmentation function are associated with the
function $S_D$, we expect
\begin{align}
\label{eq:expect}
f(m_t,\mu)\stackrel{?}{=}C_D(m_t,\mu)\,, 
\end{align}
which would imply a simple relation between the massive and massless
soft functions in (\ref{eq:softfactmassless}) through a double
convolution with partonic shape functions. We have checked that
(\ref{eq:softfactmassless}) and (\ref{eq:expect}) are satisfied for
the NLO functions, and for the part of the NNLO functions determined
by approximate NNLO formulas. In addition, we can use
(\ref{eq:expect}) as a consistency check between various NNLO
calculations available in the literature, namely those for the
renormalization factors in (\ref{eq:Zmatching}), that for the
fragmentation function, and that for the shape function. We find
nearly total agreement, with the exception of a piece related to the
$C_A C_F$ color factor. In particular, we find that direct evaluation
of (\ref{eq:Zmatching}) to NNLO yields
\begin{align}
  f(m_t,\mu)= C_D(m_t,\mu)- 4\pi^2 C_A
  C_F\left(\frac{\alpha_s}{4\pi}\right)^2 \,.
\end{align} 
Unfortunately, we have not been able to resolve the source of this
discrepancy.

\section{Convolutions of plus distributions}
\label{sec:convolutions}

Let us define the function
\begin{equation}
  f(z,\eta) \equiv \frac{e^{-2 \eta \gamma_E}}{2 \Gamma(2 \eta)} 
  \frac{1}{(1-z)^{1-2 \eta}} \, .
\end{equation}
Representations of plus distributions can be obtained by taking
derivatives of $f$ with respect to $\eta$ and by subsequently
expanding the result in the limit $\eta \to 0$. One finds for example,
with a test function $g$,
\begin{align}
  \int_0^1 dz \delta(1-z) g(z) &= \left. 2 \int_0^1 dz f(z,\eta)
    g(z)\right|_{\eta \to 0} \, , \nonumber
  \\
  \int_0^1 dz\left[ \frac{1}{1-z}\right]_+ g(z) &=
  \left. \partial_\eta \int_0^1 dz f(z,\eta) g(z)\right|_{\eta \to 0}
  \, , \nonumber
  \\
  \int_0^1 dz\left[ \frac{\ln(1-z)}{1-z}\right]_+ g(z) &= \left.
    \frac{1}{4}\partial^2_\eta \int_0^1 dz f(z,\eta) g(z)\right|_{\eta
    \to 0} + \frac{\pi^2}{6} \left. \int_0^1 dz f(z,\eta) g(z)
  \right|_{\eta \to 0}\, .
\end{align}

These representations are useful when one wants to calculate the
convolution of plus distributions. Consider for example the following
convolution
\begin{align}
  \left[ \frac{1}{1-z}\right]_+ \otimes \left[ \frac{1}{1-z}\right]_+
  = \int_z^1 \frac{dx}{x} \left[ \frac{1}{1-x}\right]_+ \left[
    \frac{1}{1-z/x}\right]_+ \, .
\end{align}
By employing the differential representation found above one finds
\begin{align}
  \label{eq:convD0D0}
  \left[ \frac{1}{1-z}\right]_+ \otimes \left[ \frac{1}{1-z}\right]_+
  &= \left. \partial_{\eta_1} \partial_{\eta_2}\int_z^1\frac{dx}{x}
    f(x,\eta_1) f\left(\frac{z}{x},\eta_2\right) \right|_{\eta_1 \to
    0, \eta_2 \to 0} \, .
\end{align}
The integration gives
\begin{align} 
  \label{eq:intD0D0} \int_z^1\frac{dx}{x} f(x,\eta_1)
  f\left(\frac{z}{x},\eta_2\right) = \frac{e^{-2 (\eta_1+\eta_2)
      \gamma_E}}{4 \Gamma(2 \eta_1 + 2 \eta_2)} {}_2F_1 \left( 2
    \eta_1, 2\eta_2, 2 (\eta_1 +\eta_2 ), 1-z\right) (1-z)^{-1+2
    (\eta_1 +\eta_2)} \, .
\end{align} 
Since we are only interested in the terms which are singular in the $z
\to 1$ limit, for the purposes of our calculation one can actually set
${}_2F_1 \to 1$. The limits for $\eta_i \to 0$ are then considerably
easier to evaluate. To do so, one makes the following analytic
continuation in the integral in (\ref{eq:intD0D0}):
\begin{align}
  (1-z)^{-1 +2 (\eta_1 +\eta_2) } \rightarrow (1-z)^{-1 +2 (\eta_1
    +\eta_2) } + \frac{\delta(1-z)}{2 (\eta_1 +\eta_2)} \, .
\end{align}
One can then safely take the derivatives with respect $\eta_1$ and
$\eta_2$ in Eq.~(\ref{eq:convD0D0}) and take the limit for vanishing
$\eta_i$ to obtain
\begin{align}
  \left[ \frac{1}{1-z}\right]_+ \otimes \left[ \frac{1}{1-z}\right]_+
  = 2 \left[\frac{\ln(1-z)}{1-z} \right]_+ -\zeta_2 \delta(1-z) \, .
\end{align}

The same kind of procedure can be employed to calculate other
convolutions. For example
\begin{align}
  \left[ \frac{\ln(1-z)}{1-z}\right]_+ \otimes \left[
    \frac{1}{1-z}\right]_+ &=\left.\partial_{\eta_2}\left[\left.
        \left(\frac{\partial_{\eta_1}^2}{4} +\zeta_2
        \right)\int_z^1\frac{dx}{x} f(x,\eta_1)
        f\left(\frac{z}{x},\eta_2\right) \right|_{\eta_1\to 0}\right]
  \right|_{\eta_2 \to 0} \nonumber
  \\
  &= \frac{3}{2} \left[ \frac{\ln^2(1-z)}{1-z} \right]_+ -
  \zeta_2\left[ \frac{1}{1-z} \right]_+ + \zeta_3\delta(1-z) \, ,
  \\
  \left[ \frac{\ln(1-z)}{1-z}\right]_+\!\! \otimes \!\!\left[
    \frac{\ln(1-z)}{1-z}\right]_+\!\!\! &=
  \left.\left(\!\frac{\partial^2_{\eta_2}}{4} +\zeta_2\!
    \right)\!\!\left[\!\left. \left(\frac{\partial_{\eta_1}^2}{4}
          +\zeta_2 \right)\!\!\int_z^1\!\frac{dx}{x} f(x,\eta_1)
        f\left(\frac{z}{x},\eta_2\right) \right|_{\eta_1\to
        0}\!\right] \right|_{\eta_2 \to 0}\!\!\!\! \nonumber
  \\
  &= \left[\frac{\ln^3(1-z)}{1-z} \right]_+ -2 \zeta_2
  \left[\frac{\ln(1-z)}{1-z} \right]_+ +2 \zeta_3\!
  \left[\frac{1}{1-z} \right]_+\!\!\! -\frac{\zeta(4)}{4} .
\end{align}

\section{Anomalous dimensions}
\label{app:AnDim}

The anomalous dimension $\gamma_{\cusp}$,  introduced in (\ref{eq:qqmatrix},\ref{eq:ggmatrix}),  has the following expansion in powers of $\alpha_s$:
\begin{align}
  \label{eq:expsmallg}
  \gamma_{\cusp}(\alpha_s) = \frac{\alpha_s}{4 \pi} \left[
    \gamma^{\cusp}_0 + \left(\frac{\alpha_s}{4 \pi} \right)
    \gamma^{\cusp}_1 + \left(\frac{\alpha_s}{4 \pi} \right)^2
    \gamma^{\cusp}_2 + {\mathcal O}(\alpha_s^3) \right] \, .
\end{align}
Completely analogous expansions hold for $\gamma^q$, $\gamma^g$,
$\gamma^S$, and $\gamma^\phi$. The coefficients of the expansion in
(\ref{eq:expsmallg}) are \cite{Moch:2004pa}
\begin{align}
  \gamma_0^{\rm cusp} &= 4 \,, \nonumber
  \\
  \gamma_1^{\rm cusp} &= \left( \frac{268}{9} - \frac{4\pi^2}{3}
  \right) C_A - \frac{80}{9}\,T_F n_f \,, \nonumber
  \\
  \gamma_2^{\rm cusp} &= C_A^2 \left( \frac{490}{3} -
    \frac{536\pi^2}{27} + \frac{44\pi^4}{45} + \frac{88}{3}\,\zeta_3
  \right) + C_A T_F n_f \left( - \frac{1672}{27} + \frac{160\pi^2}{27}
    - \frac{224}{3}\,\zeta_3 \right) \nonumber
  \\
  & \mbox{}+ C_F T_F n_f \left( - \frac{220}{3} + 64\zeta_3 \right) -
  \frac{64}{27}\,T_F^2 n_f^2 \,.
\end{align}
The coefficients in the expansion of $\gamma^q$ and $\gamma^g$ up to
${\mathcal O}(\alpha_s^2)$ are \cite{Moch:2005id, Becher:2006mr}
\begin{align}
  \gamma_0^q &= -3 C_F \,, \nonumber
  \\
  \gamma_1^q &= C_F^2 \left( -\frac{3}{2} + 2\pi^2 - 24\zeta_3 \right)
  + C_F C_A \left( - \frac{961}{54} - \frac{11\pi^2}{6} + 26\zeta_3
  \right) + C_F T_F n_f \left( \frac{130}{27} + \frac{2\pi^2}{3}
  \right) ,
\end{align}
and \cite{Moch:2005id, Becher:2007ty}
\begin{align}
  \gamma_0^g &= - \frac{11}{3}\,C_A + \frac{4}{3}\,T_F n_f \,,
  \nonumber
  \\
  \gamma_1^g &= C_A^2 \left( -\frac{692}{27} + \frac{11\pi^2}{18} +
    2\zeta_3 \right) + C_A T_F n_f \left( \frac{256}{27} -
    \frac{2\pi^2}{9} \right) + 4 C_F T_F n_f \,.
\end{align}
The coefficients in the expansion of $\gamma^S$ up to ${\mathcal
  O}(\alpha_s^2)$ are \cite{Gardi:2005yi,
  Becher:2005pd,Neubert:2007je}
\begin{align}
  \gamma^S_0 &= -2 C_F \, ,\nonumber
  \\
  \gamma^S_1 &= C_F \left[ \left( \frac{110}{27} + \frac{\pi^2}{18} -
      18 \zeta_3 \right) C_A + \left( \frac{8}{27} + \frac{2}{9}
      \pi^2\right) T_F n_f \right] \, .
\end{align}
The coefficients in the expansion of PDF anomalous dimensions up to  ${\mathcal O}(\alpha_s^2)$ are
\begin{align}
  \gamma^{\phi_q}_0 &= 3 C_F \, , \nonumber
  \\
  \gamma^{\phi_q}_1 &= C_F^2 \left( \frac{3}{2} \!-\! 2 \pi^2 \!+\!24
    \zeta_3 \right) \!+\! C_F C_A \left(\frac{17}{6} \!+\! \frac{22
      \pi^2}{9} \!-\! 12 \zeta_3 \right) \!-\! C_F T_F n_f
  \left(\frac{2}{3} \!+\! \frac{8 \pi^2}{9} \right) \, ,
\end{align}
and 
\begin{align}
  \gamma^{\phi_g}_0 &= \frac{11}{3}C_A - \frac{4}{3}T_F n_f \, ,
  \nonumber
  \\
  \gamma^{\phi_g}_1 &= C_A^2 \left( \frac{32}{3} + 12 \zeta_3 \right)
  - \frac{16}{3} C_A T_F n_f - 4 C_F T_F n_f \, .
\end{align}
for the gluon and quark PDFs respectively.

Finally, we define expansion coefficients for the QCD $\beta$ function
as
\begin{align}
  \beta(\alpha_s) &= -2\alpha_s \left[ \beta_0 \,
    \frac{\alpha_s}{4\pi} + \beta_1 \left( \frac{\alpha_s}{4\pi}
    \right)^2 + \beta_2 \left( \frac{\alpha_s}{4\pi} \right)^3 +
    \ldots \right]\, ,
\end{align}
where to three-loop order we have
\begin{align}
  \beta_0 &= \frac{11}{3} C_A - \frac{4}{3} T_F n_f \, , \nonumber
  \\
  \beta_1 &= \frac{34}{3} C_A^2 - \frac{20}{3} C_A T_F n_f - 4 C_F T_F
  n_f \, , \nonumber
  \\
  \beta_2 &= \frac{2857}{54} C_A^3 + \left( 2 C_F^2 - \frac{205}{9}
    C_F C_A - \frac{1415}{27} C_A^2 \right) T_F n_f + \left(
    \frac{44}{9} C_F + \frac{158}{27} C_A \right) T_F^2 n_f^2 \, .
\end{align}

\end{document}